\documentclass[9pt,twocolumn,twoside]{pnas-new}
\usepackage{array}
\usepackage[normalem]{ulem}
\usepackage{pdfpages}

\templatetype{pnasresearcharticle} % Choose template 
\title{Mechanistic framework for reduced-order models in soft materials: Application to \textcolor{black}{three-dimensional} granular intrusion }

\author[a]{Shashank Agarwal}
\author[b]{Daniel I Goldman}
\author[a]{Ken Kamrin}

\affil[a]{Department of Mechanical Engineering, Massachusetts Institute of Technology, Cambridge}
\affil[b]{Department of Physics, Georgia Institute of Technology, Atlanta}

\leadauthor{Agarwal} 

\significancestatement{This work proposes a general theory for modeling diverse granular intrusion problems such as animal and human locomotion in sands and other natural terrains. Respecting numerous physical constraints, the theory allows for modeling arbitrary motion of three-dimensional generally-shaped objects in granular media in near real-time. Moreover, the work provides a generic mechanistic framework for developing self-consistent physics-informed reduced-order models for a wider class of soft materials.}

\correspondingauthor{\textsuperscript{2}To whom correspondence should be addressed. E-mail: kkamrin@mit.edu}

\begin{abstract}

Soft materials often display complex behaviors that transition through apparent solid- and fluid-like regimes. While a growing number of microscale simulation methods exist for these materials, reduced-order models that encapsulate the global-scale physics are often desired to predict how external bodies interact with soft media, as occurs in diverse situations from impact and penetration problems to locomotion over natural terrains.  This work proposes a systematic program to develop three-dimensional reduced-order models for soft materials from a fundamental basis using continuum symmetries and rheological principles. In particular, we derive a reduced-order technique for modeling intrusion in granular media which we term three-dimensional Resistive Force Theory (3D-RFT), which is capable of accurately and quickly predicting the resistive stress distribution on arbitrary-shaped intruding bodies. Aided by a continuum description of the granular medium, a comprehensive set of spatial symmetry constraints, and a limited amount of reference data, we develop a self-consistent and accurate 3D-RFT. We verify the model capabilities in a wide range of cases and show it can be quickly recalibrated to different media and intruder surface types. The premises leading to 3D-RFT anticipate application to other soft materials with strongly hyperlocalized intrusion behavior.

\end{abstract}

\keywords{Soft matter $|$ Intrusion modeling $|$ Resistive Force Theory $|$ Continuum modeling $|$ Granular media \\
} 

\dates{This manuscript was compiled on \today}
\doi{\url{www.pnas.org/cgi/doi/10.1073/pnas.XXXXXXXXXX}}

\begin{document}

\maketitle
\thispagestyle{firststyle}
\ifthenelse{\boolean{shortarticle}}{\ifthenelse{\boolean{singlecolumn}}{\abscontentformatted}{\abscontent}}{}

Intrusion in soft media is a common occurrence in nature arising in biological and vehicular locomotion, excavation and anchoring applications, and meteorite and ballistic impact problems \cite{thoroddsen2001granular,agarwal2019modeling,schiebel2020mitigating,artemieva2004launch}. Modeling intrusion in real-time is critical for a variety of applications and would enable  heuristic understanding and quick insight into phenomena like biological circummutation \cite{treers2021granular} and robot-terrain interactions \cite{ijspeert2014biorobotics}. But the multiphase nature of these materials --- simultaneous solid- and fluid-like behaviors  \cite{van2017impact} --- makes modeling such systems computationally challenging. In the specific case of granular media, despite over a century of progress in the disciplines of granular physics and \textit{terramechanics}  --- the study of the interaction of tracked vehicles on various substrates \cite{he2019review}  --- challenges remain. Many commonly used methods have limited applicability due to their shape- or media-specific nature. For instance, commonly used {terramechanical} empirical models such as the Bekker model \cite{bekker1969introduction} (later modified by Wong and Reece \cite{wong1967prediction}) and Magic formulae \cite{pacejka1992magic} are limited to specific geometries such as circular wheels. Inspired by an analogous approach for viscous fluids \cite{gray1955propulsion,brokaw2006flagellar}, in recent years a granular Resistive Force Theory (RFT) has been introduced \cite{li2013terradynamics} to model the forces on arbitrarily shaped intruders in granular media, but its form is limited to 2-dimensional problems. This poses limits on its usage in many practical applications. {Attempts to extend RFT to 3D intruders have only recently been explored based on empirical fitting, though these approaches have known limitations \cite{huang2022dynamic, treers2021granular}  (see Sec S1 of Supplemental Information for comparison and critique)}.

{While granular intrusions represent a wide class of intrusion problems, equally plentiful problems exist in other soft material systems such as muds and slurries. The challenges are further exacerbated by the 3-dimensional nature of such problems that require additional physical self-consistency constraints. Thus, this work introduces a generic program for developing intrusion models in a wide class of soft materials and exemplifies its use in the case of granular media. The basic program is to combine three ingredients from the full-field physics of the soft  media to extract a ``hyper-localized'' rule-set  for determining intrusive stresses.  First, a continuum model that parsimoniously represents the rheology of the media is identified.  Second, dimensional analysis of the continuum system together with surface-media boundary stress constraints are used to obtain a generic functional form for the local intrusion stress formula.  Third, global symmetries are enforced to reduce the remaining functional dependences. In our application to granular intrusions herein, the final step is to fill in the remaining details of the resulting functions using a targetted set of in-silico reference tests. We use this program to develop a 3D-RFT model with additional efforts to keep its structure similar to the previous 2D-RFT. We test the 3D-RFT model against a variety of granular intrusions, consisting of the arbitrary motion of many symmetric and asymmetric shapes in beds of granular media. We find excellent agreement between the reference results and 3D-RFT predictions both globally (total intrusion force and moment) and locally (surface stress distribution). 
%In regards to granular materials, this work extends the work of 2-dimensional granular resistive force theory (2D-RFT) proposed by Li et al. \cite{li2013terradynamics} to the general 3-dimensional case. 
%Attempts to extend RFT to 3D intruders have only recently been explored based on empirical fitting, though these approaches have known limitations \cite{huang2022dynamic, treers2021granular}  (see Sec 1 of Supplemental Information for comparison and critique). 
Thus, the proposed set of steps, which could also be extended to other soft flowable materials, helps us develop a 3D-RFT framework that satisfies all needed physical contraints, is robust and predictive, and whose dependence on material parameters  and surface roughnesses is transparent. \textcolor{black}{We further discuss the approach's use in material systems other than non-cohesive granular media in the conclusion section and Sec S2 of the Supplementary Information.}
%Here, we propose a robust and predictive 3D-RFT framework that is easy to use and quick to recalibrate for a variety of granular media and surface roughnesses.  Moreover, it is derived under mechanistic presmises that could be extended to other soft flowable materials.  
}

\section{Review of existing RFT}
%\subsection*{Background --- Generic RFT form}
%\textcolor{black}{Our proposed model builds upon an existing 2D granular RFT form. Thus we start the process by first discussing the existing RFT form for some inspiration.} 
%The process Motivation 
The Resistive Force Theory methodology was originally introduced by Gray and Hancock \cite{gray1955propulsion} for modeling self-propelling undulatory biological systems in viscous fluids. In this model, a simple approximate formula for the resistive force on a segment of a thin body is derived from the Stokes equations as a function of the segment's velocity components, orientation, and a few variables characterizing the fluid-segment interaction. Importantly, the theory assumes decoupling of the forces over the various segments of the body \cite{brokaw2006flagellar}. The success of fluid RFT motivated multiple studies \cite{maladen2009undulatory, zhang2014effectiveness, li2013terradynamics} to explore the existence of a similar theory in granular media. 

\begin{figure}[ht!]
\centering
\includegraphics[trim = 25mm 0mm 55mm 0mm, clip, width=1.0 \linewidth] {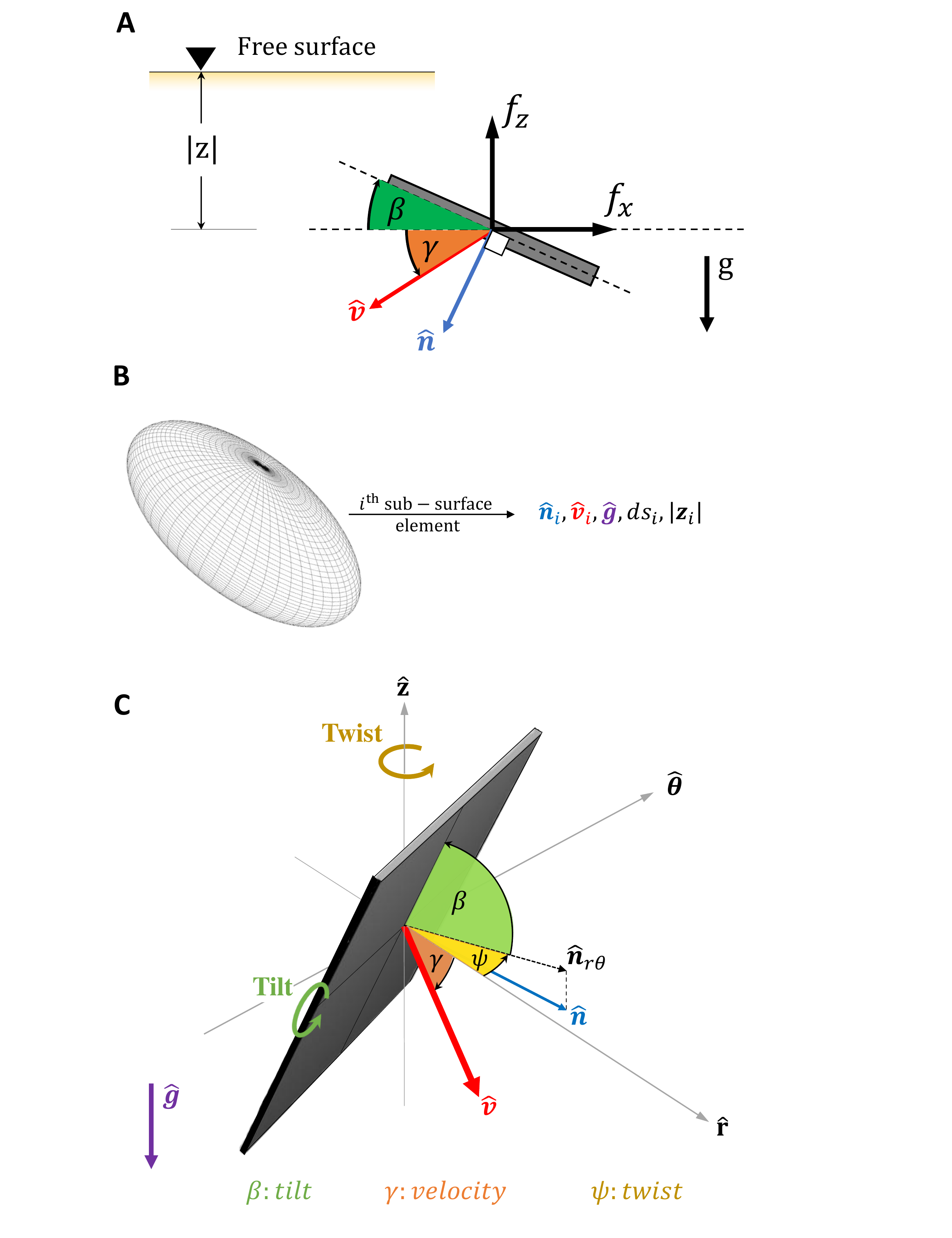}
\caption{\textcolor{black}{\emph{ 2D and 3D RFT surface characterization}: \textbf{2D-RFT --- } (A) Any moving sub-surface (line element) is represented using a set of two characteristic angles --- plate tilt ($\beta$, green) and angle of attack ($\gamma$, orange). \textbf{3D-RFT --- } (B) Any moving sub-surface (plate element) is represented using surface normal ($\hat{\boldsymbol{n}}$), area magnitude ($ds$), depth ($|{z}|$), velocity direction ($\hat{\boldsymbol{v}}$), and gravity direction ($\hat{\boldsymbol{g}}$). (C) Directions $\hat{\boldsymbol{n}}$ and $\hat{\boldsymbol{v}}$ are expressed using three characteristic angles --- plate tilt ($\beta$, green), plate twist ($\psi$, yellow), and velocity angle ($\gamma$, orange) in the local coordinate frame $\{\hat{\boldsymbol{r}},\hat{\boldsymbol{\theta}},\hat{\boldsymbol{z}}\}$}. %The functional forms of these angles and local coordinate frame are given in  Sec S2 of the Supplementary Information. 
}
\label{fig:2d_3drft_angles}
\end{figure}

Li et al \cite{li2013terradynamics} proposed a planar (or 2-dimensional) version of RFT for dry granular media (2D-RFT). In 2D-RFT, at low-speeds, the rate-independent nature of granular media (characterized by low values of the non-dimensional {(micro-) inertial number $I$} \cite{jop2006constitutive,midi2004dense,gravish2014plow} and macro-inertial number $I_{\text{mac}}$, see Materials and Methods) makes the intrusion force independent of the velocity magnitude. Assuming material strength increases with pressure and that pressure is primarily due to gravity, Li's 2D-RFT model has the following form:
\begin{align}
\boldsymbol{F}^{\text{total}} = \int_{\text{surf}} \left(\alpha_x(\beta,\gamma),\alpha_z(\beta,\gamma)\right) |z|\, ds\, . 
\label{eq:2drftbg}
\end{align} 
 Here, $\boldsymbol{F}^{\text{total}}$ represents the total force on an intruding surface, which is divided into smaller planar sub-surface elements of area $ds$ and depth $|{z}|$ from the free surface. The \emph{tilt angle} $\beta$ and \emph{angle of attack} $\gamma$ characterize the orientation and motion of each surface element of the intruding body (see Fig \ref{fig:2d_3drft_angles})A. The vector-valued function of angles $\boldsymbol{\alpha}=(\alpha_x, \alpha_z)$ represents the force per unit area per unit depth; this function must be obtained a priori through experiments or simulations of plate drag and depends on the material properties of the granular media, the intruder surface interaction, and the value of gravity. Of note, Eq \ref{eq:2drftbg} assumes no correlation between the forces on different sub-surfaces; only details local to a surface element determine the force on that element \cite{maladen2009undulatory}. A comprehensive comparison of various existing reduced order methods for modeling granular intrusions, including 2D-RFT and a \textit{terramechanical} model, can be referred from Agarwal et al. \cite{agarwal2019modeling}. 
% \begin{figure}[t]
% \centering
% \includegraphics[trim = 0mm 190mm 60mm 0mm, clip, width=1.0 \linewidth] {graphics/2DRFT_characterization.pdf}
% \caption{\emph{ 2D RFT sub-surface characterization}: Any moving sub-surface is represented using a set of two characteristic angles --- plate tilt ($\beta$, green) and angle of attack ($\gamma$, orange).}
% \label{fig:2drft_angles}
% \end{figure}

% 2d_3d_characterize_figure

In recent years, it has been shown that plasticity-based PDE models can also obtain the form of 2D granular RFT \cite{askari2016intrusion}. More recently, the performance of the continuum approach in modeling a variety of granular intrusions has been demonstrated for wheeled locomotion, impact and penetration, and multi-body intrusion \cite{dunatunga2015continuum, dunatunga2017continuum, agarwal2021efficacy,agarwal2019modeling,agarwal2021surprising}. Additionally, the approach also provides insight into the somewhat surprising observation that granular RFT is often more accurate than its viscous fluid counterpart \cite{gray1955propulsion,brokaw2006flagellar}. Thus, while experimental observations primarily drove the original RFT discoveries, the availability of faster computational methods, the success of 2D-RFT, and a need for better real-time 3D granular intrusion methods have driven the exploration of 3D-RFT. Our work combines the capabilities of the continuum approach with {a few} symmetry requirements and DEM data to accurately and efficiently model the physics of 3-dimensional granular intrusion to develop a 3D-RFT based on our proposed mechanistic framework. We briefly discuss the details of the continuum approach next.

\section{Guidance from continuum modeling} \label{sec:data_collection}
We use continuum modeling as the primary theoretical motivator as well as a reference data generation tool in this work. The constitutive model we use \cite{dunatunga2015continuum,dunatunga2017continuum} has a response characterized by a rate-insensitive, non-dilatant frictional flow rule when in a dense state, but also models the separated state which allows material to become stress-free when below a \textit{critical density}. The model has been validated in a number of previous studies of granular intrusion and locomotion \cite{askari2016intrusion,agarwal2019development,slonaker2017general,agarwal2021surprising,zhang2020expanded}. 

The constitutive flow equations representing the material's separation behavior, shear yield condition, and tensorial co-directionality, respectively, are shown below:
\begin{align}
&(\rho-\rho_c)P=0 \quad \quad \,\,\text{and} \quad P\ge 0 \quad \text{and} \quad \rho\le\rho_c, \nonumber \\
&\dot{\gamma}\, ({\tau}-\mu_{\text{int}} P)=0 \quad \text{and} \quad \dot{\gamma}\ge 0 \quad \text{and}\, \quad {\tau}\leq\mu_{\text{int}}P, \nonumber \\
&D_{ij}/\dot{\gamma} = \sigma_{ij}'/2{\tau} \quad \,\,\, \text{if} \quad \dot{\gamma}>0 \quad \text{and} \quad P>0 \label{eq:3} 
\end{align}
where \textcolor{black}{subscripts  $i,j=1,2,3$}. In these equations, $\boldsymbol{\sigma}$ represents the Cauchy stress tensor and ${\sigma}'_{ij}={\sigma}_{ij}$ + $P\delta_{ij}$ represents the deviatoric part of $\boldsymbol{\sigma}$ where $P = -{\sigma}_{ii}/3$ represents the hydrostatic pressure \textcolor{black}{(with summation implied over repeated indices)}.  ${\tau} =\sqrt{\sigma'_{ij}\sigma'_{ij}/2}$ represents the equivalent shear stress and $\mu_\text{int}$ and $\rho_c$ represent the constant bulk friction coefficient and critical close-packed density of the granular media. $D_{ij} =(\partial_iv_j+\partial_j v_i)/2$ represents the (plastic) flow rate tensor, and $\dot{\gamma}=\sqrt{2 D_{ij}D_{ij}}$ represents the equivalent shear rate. \textcolor{black}{We assume the surface friction coefficient $\mu_{\textrm{surf}}$ describes the interaction of the granular continuum with intruder surfaces. In general, $\mu_{\textrm{surf}}\le \mu_{\text{int}}$  with $\mu_{\textrm{surf}}=\mu_{\text{int}}$ in the case of a fully-rough interface.} \par

% We use the Material point method (MPM) for implementing the granular constitutive model discussed above. The authors have used this model and the method in numerous research studies in the past \cite{agarwal2019modeling,agarwal2021efficacy,agarwal2021surprising}. The implementation details can be referred from Dunatunga and Kamrin \cite{dunatunga2015continuum}. 

% Several studies in the past have verified the accuracy of this constitutive formulation in plane-strain problems. 
We use a 3D Material Point Method (MPM) solver from Baumgarten and Kamrin \cite{baumgarten2019general} to implement the continuum modeling in this study, which has been successfully used for modeling complex problems in the past \cite{baumgarten2019general,baumgarten2019general2}.  We have also validated the accuracy of the continuum model against experiments for a specific set of 3D plate intrusions, which justify the use of the continuum solver for generating 3D-RFT reference data in the final step of the model development. The details of the validation studies are provided in the Methods and Materials section.

\section{Proposed procedure: Physically-constrained intrusion modeling}\label{premises}
We begin by discussing a three-step procedure which can be used to infer reduced-order intrusion models in soft media.  This is followed by a derivation in Sec \ref{derivation} showing how these ingredients are used to deduce 3D-RFT in granular media.   \\
 
% \item \textit{Hypothesis-1: \textbf{Superposition of forces} } 
% This hypothesis assumes that the net force experienced by a large body is a vector sum of individual forces experienced by its sub-surfaces.\textbf{THIS IS NOT A HYPOTHESIS.  IT IS NEWTON'S LAW.  THIS CAN BE REMOVED. \textcolor{black}{SURE}. }
% \item \textit{Hypothesis-4: \textbf{Free-rotation about normal}} 
% This hypothesis assumes that the rotation of a plane surface along its normal axis ($\boldsymbol{\hat{n}}$) does not change the force experienced by the surface.
% \item \textit{Hypothesis-5: \textbf{Sub-surface shape independence}}
% This hypothesis assumes that the resistive forces on the plane surfaces are independent of their shapes. For instance, two surface with different shapes but with same surface normals $(\boldsymbol{\hat{n}})$, velocity directions $(\boldsymbol{\hat{v}})$, area magnitudes $(|ds|)$, and centroid depths $(|\boldsymbol{z}|)$ from the free surface, will experience same resistive forces despite having different geometric shapes. }

\noindent \textit{Step-1: \textbf{Order-reduction hypothesis.}} 
We assume that the intrusion stress on each surface element of the intruder is approximately equal to that of an isolated plate element in the same configuration moving the same way.  This is the key order-reduction hypothesis in the RFT family of models, though other reductive hypotheses could conceivably be used.\\

\noindent \textit{Step-2: \textbf{Apply constraints from continuum description.}}  The previous step reduces the problem to inferring a force relation on isolated plate elements. We now identify a continuum model for the media and use it to impose constraints on the intrusion force relation as implied by the continuum system. These constraints can be inferred through dimensional analysis of the model parameters and through analysis of stress state limitations in the rheology and boundary conditions.\\
\noindent \textit{Step-3: \textbf{Apply global symmetry constraints.}} Any function providing the intrusion force on an intruder must obey a symmetry relationship whereby if the entire problem is rotated by some amount --- that is the free-surface, gravity, intruder orientation/position, and intruder velocity are all rotated the same amount --- then the resistive force must also rotate by this common global rotation. As we will show, this constraint, which implies the drag force relations are isotropic functions of their inputs, imposes a rather strong restriction on the three-dimensional form that 3D-RFT can take. \\

It is only after Steps 1-3 have reduced down the functional form of the intrusion model considerably that we then refer to data to fit the remaining details.  Of key importance, much less fitting must be done and one is assured the result obeys basic physical principles when the above procedure is used. As we shall show with 3D-RFT, this procedure results in an accurate model with an explicit dependence on material parameters that can be exploited to enable rapid calibration to various granular media.  {\color{black} Beyond granular media, in \cite{askari2016intrusion}, a pseudo-diagnostic test was proposed to determine when a constitutive model for a material is likely to give rise to an accurate RFT-like intrusion model (Step 1).  This ``garden hoe test'' examines the mathematical form of the intrusion force under the full continuum model for the case of a finite-sized square intruder, and compares it to the scaling of intrusion force necessarily implied by the corresponding RFT model.  More discussion of this test and examples of the how to use the three-step procedure in other soft materials can be found in Sec S2 of the Supplementary Information.}

In the case of granular 3D-RFT, the execution of Step 2 uses the continuum model discussed previously and summarized in Eq \ref{eq:3}. Assuming the continuum model  holds, the resistive force on a surface element depends on the same limited set of material parameters that govern the continuum model: $\rho_c$ (the critical density), $\mu_{\text{\text{int}}}$ (the internal friction), and $\mu_{\text{surf}}$ (the media-surface friction). This requirement is quite constraining when combined with dimensional analysis.  Also, the continuum model's lack of tensile stress states is enforced by requiring resistive stress to only have positive compressive normal component and to occur only on \textit{leading edges} of the intruder. That is, only surfaces moving `into' and not `away from' a granular volume experience non-negligible resistive force. Section S3 in the Supplementary Information provides evidence in support of this hypothesis in three dimensions.  The continuum model's boundary conditions also come to use. Since the intruder is assumed to have a surface-media friction coefficient $\mu_{\text{surf}}$, the ratio of tangential and normal stress on a surface element cannot exceed this value.  In agreement with this requirement, we observe in extensive analysis of continuum model solutions that plate-tangential resistive forces generated at a higher $\mu_{\text{surf}}$ can be used to generate the tangential force for a lower $\mu_{\text{surf}}$ by simply limiting the magnitude of the tangential force based on the Coulomb friction limit.  Detailed material response graphs in this regard can be found in Sec S4-S5 of the Supplementary Information. Our extensive data analysis also allows us to assert that the normal force is relatively uninfluenced by $\mu_{\text{surf}}$ for a large range of internal friction ($\mu_{\textrm{int}} = 0.3 - 0.9$) (see Fig S3). Between different $\mu_{\textrm{int}}$, the normal forces appear to only vary by a multiplicative scalar factor $\xi_n$ as discussed in Sec \ref{derivation}.

% \textit{Premise-7: \textbf{Consistency with lower-dimensional RFT.}}  We desire a 3D-RFT model that collapses back to the previously defined 2D-RFT description in the appropriate limits. {Thus, in line with the angle-based characterization of 2D-RFT by Li et al.\cite{li2013terradynamics} (Fig \ref{fig:2d_3drft_angles})A, we desire to ultimately express 3D-RFT in terms of similar characteristic angles $\beta$ and $\gamma$ and a new twist angle $\psi$ representing the angle between the planes of plate normal and velocity direction with the vertical.}

In addition to these premises, we will utilize a few operational constraints. We desire a 3D-RFT model that collapses back to the previously defined 2D-RFT description in the appropriate limits. Thus, we desire to ultimately express 3D-RFT in terms of similar characteristic angles $\beta$ and $\gamma$ and a new twist angle $\psi$ representing the angle between the planes of plate normal and velocity direction with the vertical, similar to the angle-based characterization of 2D-RFT by Li et al.\cite{li2013terradynamics} (Fig \ref{fig:2d_3drft_angles})A. Also, we limit ourselves to quasi-static intruder motions, with negligible inertial effects in the granular media. This was also assumed in the original 2D-RFT formulation and lets the force on a sub-surface be deemed independent of the surface's speed. More recently, an inertia-sensitive 2D-RFT has also been proposed and validated \cite{agarwal2021surprising}. We limit our attention to quasi-static cases in this work (See Materials and Methods section for more details). 
%\textcolor{black}{ We use a combination of the granular micro- inertial number $(I)$ and a macro-inertial number $(I_{\text{mac}})$, to determine the quasi-static conditions in a granular intrusion system. Low values of $I$ and $I_{\text{mac}}$ indicate insignificant \textit{micro-inertial} and \textit{macro-inertial} effects in gravity-loaded granular intrusion systems. In general, 3D-RFT formulation can be applied to systems with $I<0.05$ and $I_{\text{mac}}<0.4$. See Materials and Methods section for definitions of these numbers and other details.} 
{\color{black}We also require that intruders are submerged to a depth $|z|$ less than a $O(10)$ factor of the size of the intruder. This requirement comes from observations that the intrusion force stops growing linearly with $|z|$ below a critical depth in  gravity-loaded quasi-semi-infinite  granular beds \cite{agarwal2021efficacy,guillard2014lift};  specifically, the lift component of the intrusion force saturates beneath the critical depth.  This interesting phenomenon occurs even though the pressure field within the grains, excluding a localized zone about the intruder, continues growing linearly with depth.} Lastly, the RFT form assumes grains to be small relative to the size-scale of the intruder. RFT is expected to have reduced accuracy along intruder surfaces that sharply vary; direct grain-size effects may be important to determining the resistive force on these subsurfaces.

\section{Deducing physically-constrained 3D-RFT}\label{derivation}
We use the previously discussed steps  to propose a general form of the intended 3D-RFT model. In light of Step 1, we propose a 3D-RFT that supposes the force on any small surface element of the intruding body is equal to what the force would be if the plate element were isolated and moving on its own.  Hence, the force (per area per depth, $\boldsymbol{\alpha}$) is a function that depends only on the element's surface normal $\hat{\boldsymbol{n}}$, local velocity direction $\hat{\boldsymbol{v}}$, and depth $|z|$, along with the acceleration of gravity $\boldsymbol{g}$ and material properties `mat', such that the total intrusion force satisfies
\begin{align}
\boldsymbol{F}^{\text{total}} = \int_{\text{surf}} \boldsymbol{\boldsymbol{\alpha}}(\hat{\boldsymbol{n}}, \hat{\boldsymbol{v}}, \boldsymbol{g}, |{z}|;\textit{mat})\, |z|\,  ds \, .
\label{eq:genericrft}
\end{align} 
Referring to Step 2, the material properties are taken to be given by the parameter set $\textit{mat}=\{\rho_c,\mu_{\text{int}},\mu_{\text{surf}}\}$. Assuming for the time being that the intruder is fully rough, $\mu_{\text{surf}}=\mu_{\text{int}}$, dimensional analysis together with the observed dependence on $\mu_{\text{int}}$ in Fig S2 reduces the functional dependence of $\boldsymbol{\alpha}$ significantly, requiring that %requires that the function %$\boldsymbol{\alpha}$ must have the form
\begin{equation}\boldsymbol{\alpha}=\rho_c g \hat{f}(\mu_{\text{int}})\,  \boldsymbol{\alpha}^{\text{gen}}(\hat{\boldsymbol{n}},\hat{\boldsymbol{v}},\hat{\boldsymbol{g}})\label{agen}
\end{equation}
where $\boldsymbol{g}=g\hat{\boldsymbol{g}}$, the dimensionless function $\hat{f}$ is as-yet undetermined, and the prefactor $\rho_c g \hat{f}(\mu_{\text{int}})$, which we collectively refer to as $\xi_n$, is a media dependent scaling coefficient reflecting the overall intrusive strength of the system.   The \textit{generic} RFT function $\boldsymbol{\alpha}^{\text{gen}}$ is labeled as such because, under the given premises, it is universal across all granular/intruder systems with fully-rough interfaces.  {\color{black}We now show how  $\boldsymbol{\alpha}^{\text{gen}}$ can be used to enable the modeling non-fully-rough surfaces.}

We can uniquely decompose the vector-valued function $\boldsymbol{\alpha}^{\text{gen}}$ into normal and tangential directions as $\boldsymbol{\alpha}^{\text{gen}}\equiv\boldsymbol{\alpha}_n^{\text{gen}}+\boldsymbol{\alpha}_t^{\text{gen}}$.  We may now remove the fully-rough assumption and suppose $\mu_{\text{surf}}\neq \mu_{\text{int}}$.  Then, in accord with the surface friction limit and conclusions drawn from Section S5, we can simply scale down the tangential component of surface stress {\color{black}from the fully-rough case} to the new surface friction {\color{black}$\mu_{\text{surf}}$} limit by writing
\begin{align}
\boldsymbol{\alpha}=& \rho_c g \hat{f}(\mu_{\text{int}})\, \left[  \boldsymbol{\alpha}_{n}^{\text{gen}}+ \text{min}\left(\frac{\mu_{\text{surf}}\, |\boldsymbol{\alpha}_{n}^{\text{gen}}|}{|{\boldsymbol{\alpha}}_{t}^{\text{gen}}|},\, 1\right)  {{\boldsymbol{\alpha}}_{t}^{\text{gen}}}\,\right]  . \label{eq:3drftformula_alpha_t}
\end{align}

The 3D-RFT model we are proposing is closed upon choosing the scalar-valued function $\hat{f}(\mu_{\text{int}})$ and the vector valued function $\boldsymbol{\alpha}^{\text{gen}}(\hat{\boldsymbol{n}},\hat{\boldsymbol{v}},\hat{\boldsymbol{g}})$.  Upon selection of these two functions, Eq \ref{eq:3drftformula_alpha_t} can be used to determine $\boldsymbol{\alpha}$ for any choice of material and interface properties $\{\rho_c,\mu_{\text{int}},\mu_{\text{surf}}\}$.

We now apply symmetry constraints inherent to the drag problem (Step 3) to further constrain the functional form of $\boldsymbol{\alpha}^{\text{gen}}$. Our strategy is to constrain the function space to satisfy symmetry constraints \textit{by design} rather than leaving it to chance based on the choice of fit functions. Moreover, by enforcing the symmetry constraints directly, we reduce the space of admissible functions, thereby reducing the amount of fitting that must be done. 

\begin{figure*}[ht!]
\centering
\includegraphics[trim = 0mm 10mm 0mm 0mm, clip, width=1.0 \linewidth] {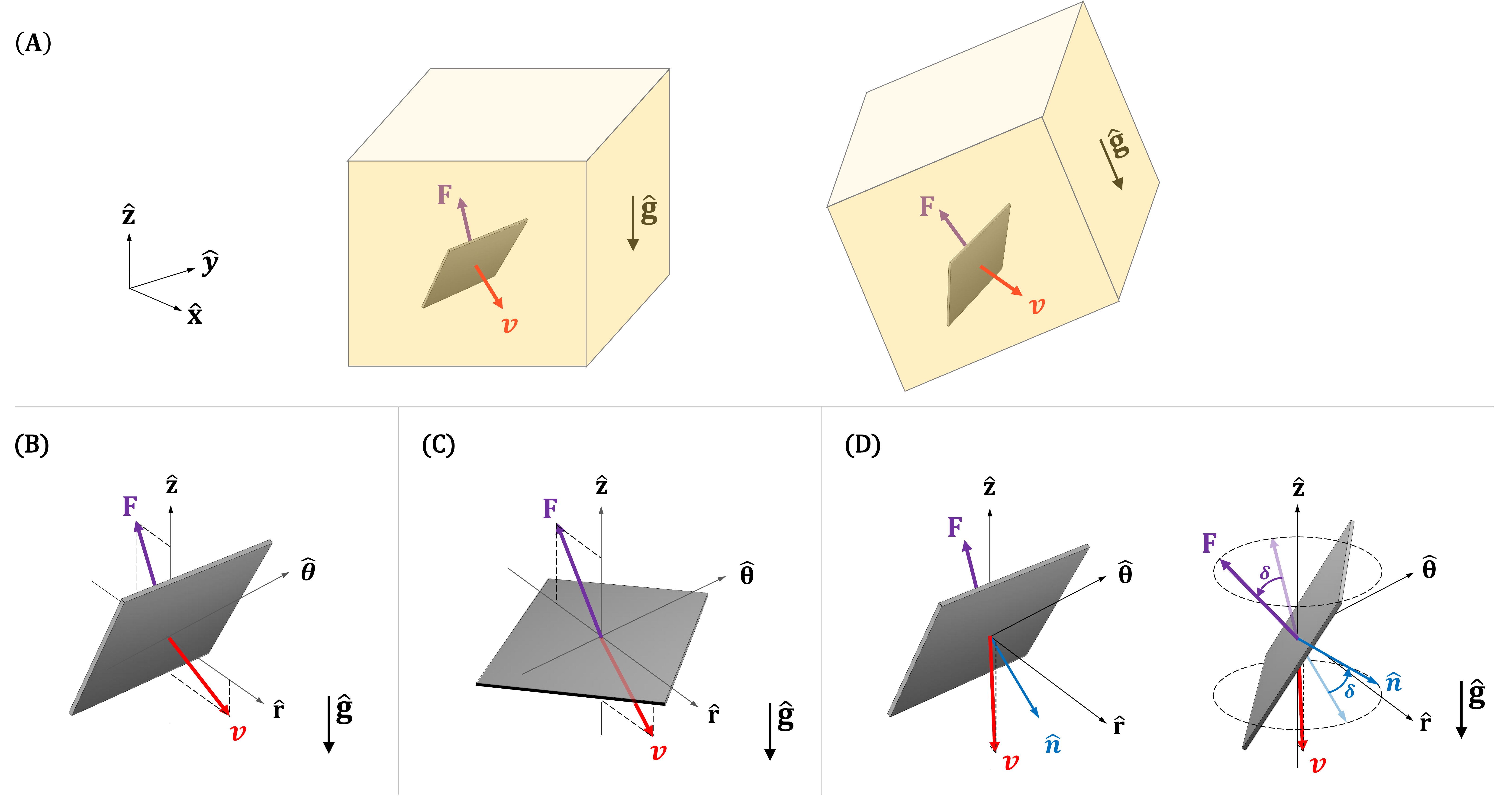}
\caption{\emph{3D-RFT symmetry constraints}: \emph{(A)} \textit{Global rotational constraint} requiring the drag force to be an isotropic function of the plate normal, motion direction, and gravity direction.  Some consequences of this constraint are  \textit{plate twist symmetry}, \textit{plate tilt symmetry}, and \textit{vertical motion symmtry}. \emph{(B)} A special case of \textit{plate twist symmetry:} $F_\theta(\beta,\gamma,\psi=0)=0$. \emph{(C)} A special case of \textit{plate tilt symmetry:} $F_{\theta}(\beta=0,\gamma,\psi) =0$, and \emph{(D)}  \textit{Vertical motion symmetry}:  $(\beta,\gamma=\pm\pi/2,\psi=0)\to (\beta,\gamma=\pm\pi/2,\psi=\delta)$ causes $(F_r,F_\theta=0,F_z) \to (F_r\cos{\delta},F_{r}\sin{\delta},F_z)$. Violet, red, and blue arrows show force, velocity, and surface-normal direction, respectively. 
%%% In the \gamma->pi/2 limit, r-\theta components are related by plate twist angles as m(\beta)\cos(\psi), m(\beta)\sin(\psi), n(\beta)
%%%% in \beta->0 \psi depenedece goes away 
}
\label{fig:all_constr}
\end{figure*}

Consider a small plate intruder characterized with $\hat{\boldsymbol{n}}$, $\hat{\boldsymbol{v}}$, $ds$, $|\boldsymbol{z}|$, and $\boldsymbol{g}$. For $\mu_{\text{surf}}=\mu_{\text{int}}$, the force on the plate according to RFT is $\boldsymbol{df}=\xi_n\boldsymbol{\alpha}^{\text{gen}}(\hat{\boldsymbol{n}},\hat{\boldsymbol{v}},{\boldsymbol{g}})|z|ds$. If the entire system is rotated --- including the intruder, the granular bed, and gravity --- the resistive force on the intruder must rotate by the same amount. This is because rotating the entire system should be consistent with a fixed system and a rotation of the observer. Figure \ref{fig:all_constr}A visualizes this action; \textcolor{black}{note the distance to the free surface along the gravity vector ($|z|$) remains unchanged as does the plate area ($ds$)}.  Thus, for any rotation $\boldsymbol{R}$, we expect that $\boldsymbol{R}\, \boldsymbol{df}=\xi_n\boldsymbol{\alpha}^{\text{gen}}(\boldsymbol{R}\hat{\boldsymbol{n}},\boldsymbol{R}\hat{\boldsymbol{v}},\boldsymbol{R}\hat{\boldsymbol{g}})|z|\, ds$, and thus
%We consider that this intruder is a square plate of unit side (and thus unit area) and is submerged to a unit depth. This results in forces experienced by this intruder to be equal to $\boldsymbol{\alpha}$. \\
\begin{align}
\boldsymbol{\alpha}^{\text{gen}}(\boldsymbol{R}\hat{\boldsymbol{n}},\boldsymbol{R}\hat{\boldsymbol{v}},\boldsymbol{R}\hat{\boldsymbol{g}})= \boldsymbol{R}\boldsymbol{\alpha}^{\text{gen}}(\hat{\boldsymbol{n}},\hat{\boldsymbol{v}},\hat{\boldsymbol{g}})\, .
\label{eq:constr-1}
\end{align}
This \textit{`global rotation constraint'} 
implies $\boldsymbol{\alpha}^{\text{gen}}$ is an isotropic function of its inputs.  Thus, in accord with Isotropic Representation Theory (IRT)\cite{smith1971isotropic} the function must have the following specific form:
\begin{align}
 \boldsymbol{\alpha}^{\text{gen}}(\hat{\boldsymbol{n}},\hat{\boldsymbol{v}},\hat{\boldsymbol{g}}) &= f_1\hat{\boldsymbol{n}} + f_2\hat{\boldsymbol{v}}+f_3\hat{\boldsymbol{g}}\, , \label{eq:irt}
\end{align}
 where $f_1$, $f_2$, and $f_3$ are three mutually-independent arbitrary scalar-valued functions of coordinate-invariant \textit{dot-products} between the three direction vectors, that is $f_i= f_i(\hat{\boldsymbol{g}}\cdot \hat{\boldsymbol{v}},\hat{\boldsymbol{g}}\cdot \hat{\boldsymbol{n}},\hat{\boldsymbol{n}}\cdot \hat{\boldsymbol{v}})$. Equation \ref{eq:irt} has reduced the problem of fitting $\boldsymbol{\alpha}^{\text{gen}}$ from determining a vector-valued function of six independent variables (three vectors, each with a constraint of being unit magnitude) to determining a vector-valued function of three independent variables (three dot products). Note that the form given in Eq \ref{eq:3drftformula_alpha_t} for general $\mu_{\text{surf}}$ continues to satisfy the IRT requirement  Eq \ref{eq:irt}. A detailed proof in this regard is provided in Sec S6 of the Supplementary Information. 
 
 We next introduce the methodology for parametrizing subsurfaces in terms of three angles to arrive at our ultimate description of $\boldsymbol{\alpha}^{\text{gen}}$. 

 %Our proposed form of generic 3D-RFT expresses $f_{\mu_{\text{surf}}}$ in $|\boldsymbol{\alpha}_t|$ at a high $\mu_{\text{surf}}$. The corresponding values of $|\boldsymbol{\alpha}_t|$ for any $\mu_{\text{surf}}$ can be obtained from 3D-RFT generic values by limiting the $|\boldsymbol{\alpha}_t|$ to $\mu_{\text{surf}}|\boldsymbol{\alpha}_n|$. Thus, $f_{\mu_{\text{surf}}}$ is never explicitly used during 3D-RFT implementation. More details are discussed in later sections. We now discuss various assumptions (and their justifications) used for developing this form of 3D-RFT.

\subsection*{3D-RFT sub-surface characterization} 

Equation \ref{eq:irt} defines the normalized stress-per-depth on a sub-surface using $\hat{\boldsymbol{n}}$, $\hat{\boldsymbol{v}}$, and $\hat{\boldsymbol{g}}$ directions and corresponding dot products. We could stop here and set out to fit the $f_i$ functions, however, there are certain advantages to first re-expressing Eq \ref{eq:irt} in terms of an orthogonal set of directions and angles measured from those directions.  Using angles helps us meet our desire to maintain a consistency of 3D-RFT with the 2D-RFT form, which is also angle-based, and using an orthogonal basis rather than $\{\hat{\boldsymbol{n}},\hat{\boldsymbol{v}},\hat{\boldsymbol{g}}\}$ eases the physical interpretation and simplifies calibration. 
%Secondly, unlike $\{\hat{\boldsymbol{n}},\hat{\boldsymbol{v}},\hat{\boldsymbol{g}}\}$, $\{\hat{\boldsymbol{r}},\hat{\boldsymbol{\theta}},\hat{\boldsymbol{z}}\}$  represents an orthogonal co-ordinate frame, which eases the physical interpretation of the componenets of $\boldsymbol{\alpha}^{\text{gen}}$ and simplifies calibration.

We define a local cylindrical coordinate system at each surface element as follows (see Fig \ref{fig:2d_3drft_angles}C): We choose the direction opposite to the gravity (upward in general) as the positive $z$-direction and use the horizontal component of $\hat{\boldsymbol{v}}$ as the positive $\hat{\boldsymbol{r}}$ direction. The remaining $\hat{\boldsymbol{\theta}}$ direction is chosen as the cross product between $\hat{\boldsymbol{r}}$ and $\hat{\boldsymbol{z}}$. The free-surface is taken as the reference ($z=0$) for the $z$-direction.\footnote{When $|\boldsymbol{v}-(\boldsymbol{v}\cdot \hat{\boldsymbol{z}})\hat{\boldsymbol{z}}|$ is zero (a sub-surface moves up or down), $\hat{\boldsymbol{r}}$ is set to the direction of the horizontal component of the surface-normal i.e. $\hat{\boldsymbol{r}} = (\hat{\boldsymbol{n}} - (\hat{\boldsymbol{n}}\cdot \hat{\boldsymbol{z}}) \hat{\boldsymbol{z}})/|\hat{\boldsymbol{n}} - (\hat{\boldsymbol{n}}\cdot \hat{\boldsymbol{z}}) \hat{\boldsymbol{z}}|$.}
% \begin{figure}[ht!]
% \centering
% \includegraphics[trim = 0mm 0mm 160mm 0mm, clip, width=1.0 \linewidth] {graphics/beta_g_psi_2.pdf}
% \caption{\emph{ 3D-RFT surface element characterization}: (A) Any moving sub-surface is represented using surface normal ($\hat{\boldsymbol{n}}$), area magnitude ($ds$), depth ($|{z}|$), velocity direction ($\hat{\boldsymbol{v}}$), and gravity direction ($\hat{\boldsymbol{g}}$). (B) Directions $\hat{\boldsymbol{n}}$ and $\hat{\boldsymbol{v}}$ are expressed using three characteristic angles --- plate tilt ($\beta$, green), plate twist ($\psi$, yellow), and velocity angle ($\gamma$, orange) in the local coordinate frame $\{\hat{\boldsymbol{r}},\hat{\boldsymbol{\theta}},\hat{\boldsymbol{z}}\}$. The functional forms of these angles and local coordinate frame are given in Eq \ref{localcorrdinate_definition} and \ref{beta_eq}- \ref{psi_eq}. }
% \label{fig:3drft_angles}
% \end{figure}
Next, we recast Eq \ref{eq:irt} in terms of angles referenced against directions $\{\hat{\boldsymbol{r}},\hat{\boldsymbol{\theta}},\hat{\boldsymbol{z}}\}$.  
%two angles, the angle of twist ($\psi$) and the angle of tilt ($\beta$), to characterize the sub-surface normal vector $\hat{\boldsymbol{n}}$.  Then, the motion direction $\hat{\boldsymbol{v}}$ is characterized by the angle of attack ($\gamma$).  \\
The \textit{surface twist angle}, $\psi$, gives the azimuthal angle between the $r$-axis and the projection of the surface normal onto the $r\theta$-plane, denoted by  $\hat{\boldsymbol{n}}_{r\theta}$. The 
\textit{surface tilt angle}, $\beta$, is the polar angle between the $r$-axis and the $r\theta$-plane. To be clear, $\beta$ measures the angle between the $r\theta$-plane and one of $\boldsymbol{\hat{n}}$ or $-\boldsymbol{\hat{n}}$, whichever gives a result in the $[-\pi/2,\pi/2]$ range.
%Thus, $\beta$ definition does not differentiates between sub-surface orientations separated by $\pi$ radians in $\hat{\boldsymbol{n}}-\hat{\boldsymbol{z}}$ plane and the same value of $\beta$ applies to both $\hat{\boldsymbol{n}}$ and $-\hat{\boldsymbol{n}}$. %As a result, $\beta$ domain in 3D-RFT graphs is restricted to $[-\pi/2,\pi/2]$ instead of $[-\pi, \pi]$. 
This choice is not problematic because at any time, only one side of a plate element experiences forces, and this can be identified using the leading edge condition ($\hat{\boldsymbol{v}}\cdot \hat{\boldsymbol{n}} \ge 0$).
The local coordinate frame definitions keep the velocity vector completely within the $rz$-plane. Thus, once $\{\hat{\boldsymbol{r}},\hat{\boldsymbol{\theta}},\hat{\boldsymbol{z}}\}$ are determined, only one angle is needed to represent the velocity direction. This \textit{angle of attack}, $\gamma$, is the angle between the velocity direction vector and the local positive $r$-axis. 
% \begin{align}
% \gamma \quad = \quad  \cos^{-1}(\hat{\boldsymbol{v}} \cdot  \hat{\boldsymbol{r}}) \quad & \mathrm{if} \quad \hat{\boldsymbol{v}}\cdot \hat{\boldsymbol{z}}\le 0 \nonumber \\
%         \quad - \cos^{-1}(\hat{\boldsymbol{v}} \cdot  \hat{\boldsymbol{r}}) \quad & \mathrm{if} \quad \hat{\boldsymbol{v}}\cdot \hat{\boldsymbol{z}} \ge 0 \label{gamma_eq}
% \end{align}
%We interchangeably use $\gamma_{vel}$ and $\gamma$ in this text for a better understanding of readers, but they both correspond to the same angle definition above. 
See Fig \ref{fig:2d_3drft_angles}C for a visual representation of these angles. Based on the above definitions, the variations of each of the system characteristic angles $\{\beta,\gamma,\psi\}$ is restricted to $[-\pi/2,\pi/2]$ for any leading-edge surface. We use these limits in the generation of reference 3D-RFT data. Mathematical formulae for the angles in terms of vector components in a fixed cartesian frame are provided in Sec S7 of the Supplementary Information.
% \begin{align}
% \beta \in \big[-\pi/2,\pi/2\big] \nonumber \\
% \gamma \in \big[-\pi/2,\pi/2\big] \nonumber \\
% \psi \in \big[-\pi/2,\pi/2\big]
% \end{align}
With some algebra, one can express the $\{\hat{\boldsymbol{n}},\hat{\boldsymbol{v}},\hat{\boldsymbol{g}}\}$ basis vectors in terms of $\{\hat{\boldsymbol{r}},\hat{\boldsymbol{\theta}},\hat{\boldsymbol{z}}\}$ and the three angles (see Eq 9 of Supplementary Information). 
Substituting the result into Eq \ref{eq:irt} yields the expressions for the components of $\boldsymbol{\alpha}^{\text{gen}}=\alpha_r^{\text{gen}}\hat{\boldsymbol{r}} + \alpha_{\theta}^{\text{gen}}\hat{\boldsymbol{\theta}} + \alpha_z^{\text{gen}}\hat{\boldsymbol{z}}$ as follows: 
% We express the final form of $\boldsymbol{\alpha}^{\text{gen}}$ in the local coordinate frame $\{\hat{\boldsymbol{r}},\hat{\boldsymbol{\theta}},\hat{\boldsymbol{z}}\}$ by expressing  $\{\hat{\boldsymbol{n}},\hat{\boldsymbol{v}},\hat{\boldsymbol{g}}\}$ as
%\begin{align}
%& \hat{\boldsymbol{g}} = -\hat{\boldsymbol{z}}\, , \qquad \qquad
%\hat{\boldsymbol{v}} = \cos{\gamma}\,\hat{\boldsymbol{r}} - \sin{\gamma}\,\hat{\boldsymbol{z}}\, , \nonumber \\
%& \hat{\boldsymbol{n}} =\sin{\beta}\cos{\psi} \,\hat{\boldsymbol{r}}  + \sin{\beta}\sin{\psi} \,\hat{\boldsymbol{\theta}} - \cos{\beta} \,\hat{\boldsymbol{z}}\, .  \label{nvg}
%\end{align}
%Substitution of definitions in Eq \ref{eq:irt} gives the expressions for the components of $\boldsymbol{\alpha}^{\text{gen}}=\alpha_r^{\text{gen}}\hat{\boldsymbol{r}} + \alpha_{\theta}^{\text{gen}}\hat{\boldsymbol{\theta}} + \alpha_z^{\text{gen}}\hat{\boldsymbol{z}}$ as follows: 
\begin{align}
%\boldsymbol{\alpha}^{\text{gen}}
%(\hat{\boldsymbol{n}},\hat{\boldsymbol{v}},\hat{\boldsymbol{g}}) 
%&= \alpha_r^{\text{gen}}\hat{\boldsymbol{r}} + \alpha_{\theta}^{\text{gen}}\hat{\boldsymbol{\theta}} + \alpha_z^{\text{gen}}\hat{\boldsymbol{z}} \label{eq:alpha_in r_theta_z} \\
\alpha_r^{\text{gen}}(\beta,\gamma,\psi)       &= f_1 \sin{\beta} \cos{\psi} + f_2 \cos{\gamma} \nonumber \\
\alpha^{\text{gen}}_{\theta}(\beta,\gamma,\psi)&= f_1 \sin{\beta} \sin{\psi} \nonumber \\
\alpha^{\text{gen}}_z(\beta,\gamma,\psi)&=-f_1 \cos{\beta} - f_2 \sin{\gamma} -f_3 
\label{3drft_eq}
\end{align}
%where, $f_{1}=f_1(x_1,x_2,x_3)$, $f_{2}=f_2(x_1,x_2,x_3)$, and $f_{3}=f_3(x_1,x_2,x_3)$ 
where $f_i=f_i\left(  \hat{\boldsymbol{g}}\cdot \hat{\boldsymbol{v}} ,\, \hat{\boldsymbol{g}}\cdot \hat{\boldsymbol{n}} ,\, \hat{\boldsymbol{n}}\cdot \hat{\boldsymbol{v}}\right)$ are three as-yet undetermined functions of the three dot products, which are now given by the three angles as follows:
%\begin{align}
%x_1 \equiv \hat{\boldsymbol{g}}\cdot \hat{\boldsymbol{v}} , \qquad  x_2 \equiv \hat{\boldsymbol{g}}\cdot \hat{\boldsymbol{n}} , \qquad x_3 \equiv \hat{\boldsymbol{n}}\cdot \hat{\boldsymbol{v}} \label{x1x2x3}
%\end{align}
%$\{x_1,x_2,x_3\}$ can be further %re-expressed in terms of $\{\beta, %\gamma, \psi\}$ using Eq \ref{nvg}.
 \begin{align}
 &\hat{\boldsymbol{g}}\cdot \hat{\boldsymbol{v}} = \sin{\gamma}\, , \qquad  \hat{\boldsymbol{g}}\cdot \hat{\boldsymbol{n}} = \cos{\beta}\, , \nonumber \\ 
 & \hat{\boldsymbol{n}}\cdot \hat{\boldsymbol{v}} = \cos{\psi}\cos{\gamma}\sin{\beta} + \sin\gamma \cos\beta\, . \label{x1x2x3}
 \end{align}
\begin{figure*}[ht!]
\centering
\includegraphics[trim = 0mm 15mm 0mm 0mm, clip, width=0.95 \linewidth] {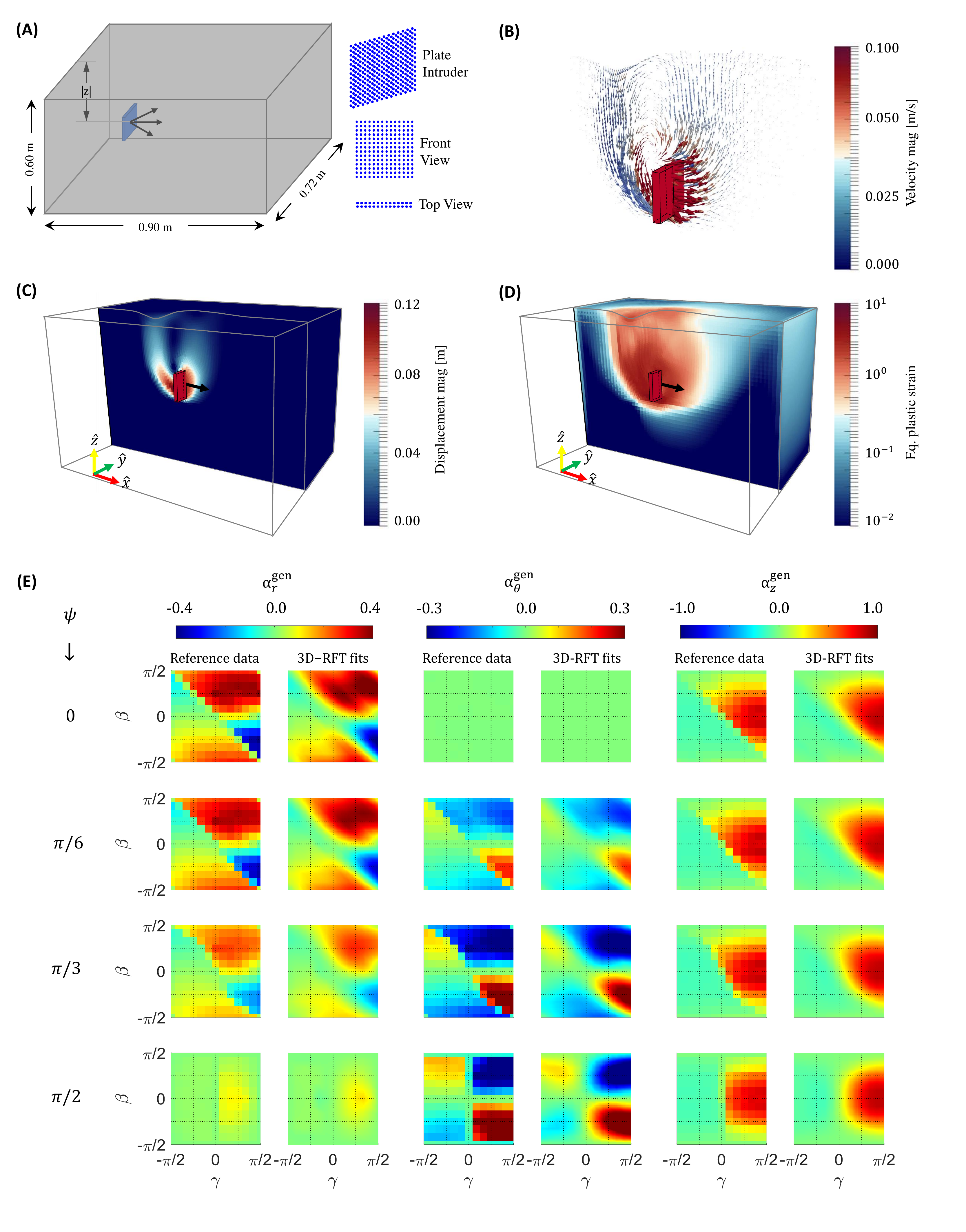}
\caption{\textcolor{black}{\emph{Reference data collection and sample 3D-RFT fitings}: We use the material point method (MPM) to simulate the continuum model for reference data collection. (A) Schematic of intrusion setup with the thin plate (0.105m×0.105m×0.015m) used for data collection. Variation of (B) material flow, (C) displacement magnitude, and (D) equivalent plastic strain magnitude from one of the test setups. %Material properties are provided in the \textit{Reference Data} section.
(E) Reference data for normalized forces ($|\boldsymbol{F}/{A|\boldsymbol{z}|\xi_n}|$) and 3D-RFT functional fittings ($|\boldsymbol{\alpha}_{r,\theta,z}^{\text{gen}}|$) for plate intrusions at various plate twists ($\psi = [0,\pi/6,\pi/3,\pi/2]$ rad), plate inclinations ($\beta= -\pi/2:\pi/6:\pi/2$ rad), and velocity directions ($\gamma = -\pi/2:\pi/6:\pi/2$ rad) for a material with $\mu_{\textrm{int}}=0.4$, $\rho=3000$ kg/m$^3$, and $\mu_{\textrm{surf}}=0.15$. The reference data is normalized with $\xi_n=0.92\times10^6$ N/m$^3$. %\textbf{[SHASHANK: I think for E it would be clearer if you also added the word ``MPM'' and ``Fit'' under each column in the figure panel.  It's hard to figure out what is meant by left and right since E has three lefts and three rights!]}
}}
\label{fig:3drftgraphs_and_setup}
\end{figure*}
Equations \ref{3drft_eq} and \ref{x1x2x3} give the final functional form of $\boldsymbol{\alpha}^{\text{gen}}$ and the completion of the three-step process outlined in Sec \ref{premises}. The 3D-RFT model we introduce is closed upon fitting the three $f_i$ as functions of the three dot products, which we shall do in the next section using a targetted set of in-silico reference tests. Note that by building the angle dependences of $\alpha_r^{\text{gen}},\ \alpha_\theta^{\text{gen}},\ \text{and}\ \alpha_z^{\text{gen}}$ indirectly from the $f_i$ using IRT rather than by directly fitting the $\alpha^{\text{gen}}$ functions, the model is guaranteed to  satisfy many {easy-to-observe} requirements regardless of how the $f_i$ are picked. These  include (i) \textit{`plate twist symmetry' }(Fig \ref{fig:all_constr}B), which requires that the sub-surface forces in the $r$- and $z$-direction should be even functions of plate twist ($\psi$), and that force in the $\theta$-direction should be an odd function of $\psi$; (ii) \textit{`plate tilt symmetry'} (Fig \ref{fig:all_constr}C) which requires that when the plate faces upwards or downwards ($\beta=0$), the sub-surface force in the $\theta$-direction should vanish, the force magnitude should depend only on $\gamma$, and the twist angle $\psi$ should have no influence on the force;  (iii) \textit{`vertical motion symmetry'} (Fig \ref{fig:all_constr}D), which requires that for any tilt $\beta$, as $\gamma\to\pm\pi/2$ (approaching an upward or downward motion) any {azimuthal rotation (changing $\psi$ at constant $\beta$)} of a sub-surface should rotate the resultant force on the sub-surface by the same angle. Moreover, by using Eqs \ref{3drft_eq}-\ref{x1x2x3} we are ensured the relation for $\boldsymbol{\alpha}^{\text{gen}}$ always has the correct periodicity in the three angles.
% \begin{figure*}[ht!]
% \centering
% \includegraphics[trim = 0mm 5mm 10mm 0mm, clip, width=0.90\linewidth] {graphics/data_collection.pdf}
% \caption{\emph{Reference data collection for 3D-RFT}: (A) We use a thin plate (0.105m×0.105m×0.015m) intrusion setup as shown in the schematic for reference data collection for 3D-RFT. We use material point method (MPM) based continuum modeling for data collection. We run 13 combinations of plate tilt angle ($\beta = -\pi/2:\pi/6:\pi/2$ rad), 13 combinations of velocity direction angle ($\gamma = -\pi/2:\pi/6:\pi/2$ rad), and 7 combinations of plate twist angle ($\psi = 0:\pi/6:\pi/2$. (B) material flow, (C) displacement magnitude, and (D) equivalent plastic strain magnitude variation from one of the test setups. Material properties are provided in the \textit{Reference Data} section.}
% \label{fig:data_collection}
% \end{figure*}

%{At this point, we have completed the three steps of our model development listed in Sec \ref{premises}. %We started with choosing an approariate representation of granular beds with continuum modeling in Section 2. We used this form to decide a general form of 3DRFT to understand the dependency on materials properties. The steps represent step-1 of our proposed model. We next use these  
%As the final step of the development, we fit the 3D-RFT form (Eqs \ref{3drft_eq} - \ref{x1x2x3}) using a set of in-silico reference tests.}}

\section{Reference data}
{\color{black}We use a large number of combinations ($\sim3000$) of material properties ($\mu_{\textrm{int}}\ \text{and} \ \mu_{\textrm{surf}}$) and 3D-RFT angles ($\beta, \gamma, \text{and} \ \psi$) to generate a bank of continuum modeling-based reference data for evaluating the 3D-RFT form. 
The details of the combinations are provided in  Sec S4 of the Supplementary Information. Based on this extensive data set, we can fit the functions ${f}_1, \, {f}_2, \, \textrm{and}\ {f}_3$ that determine $\boldsymbol{\alpha}^{\text{gen}}$ and we can also fit $\hat{f}$. } Figure \ref{fig:3drftgraphs_and_setup}A-D shows the simulation setup used for the data collection. While both the $\beta$ and the $\gamma$ angles are varied over  the interval $[-\pi/2,\pi/2]$, $\psi$ was varied only in $[0,\pi/2]$ taking advantage of \textit{`plate twist symmetry'} discussed earlier.

% \begin{figure*}[ht!]
% \centering
% \includegraphics[trim = 10mm 35mm 15mm 0mm, clip, width=0.90 \linewidth] {graphics/sample_rft_fitting.pdf}
% \caption{\emph{Sample 3D-RFT fittings}: Reference normalized forces ($F/{A|\boldsymbol{z}|\xi_n}$) and functional fittings (right) for plate intrusions at various plate twists ($\psi = [0,\pi/6,\pi/3,\pi/2]$ rad), plate inclinations ($\beta= -\pi/2:\pi/6:\pi/2$ rad), and velocity directions ($\gamma = -\pi/2:\pi/6:\pi/2$ rad) for a material with $\mu_{\textrm{int}}=0.4$, $\rho=3000$ kg/m$^3$, and $\mu_{\textrm{surf}}=0.15$. The reference data is normalized with $\xi_n=0.92\times10^6$ N/m$^3$.}
% \label{fig:3drftgraphs}
% \end{figure*}

\begin{figure*}[ht!]
\centering
\includegraphics[trim = 0mm 0mm 0mm 0mm, clip, width=0.96 \linewidth] {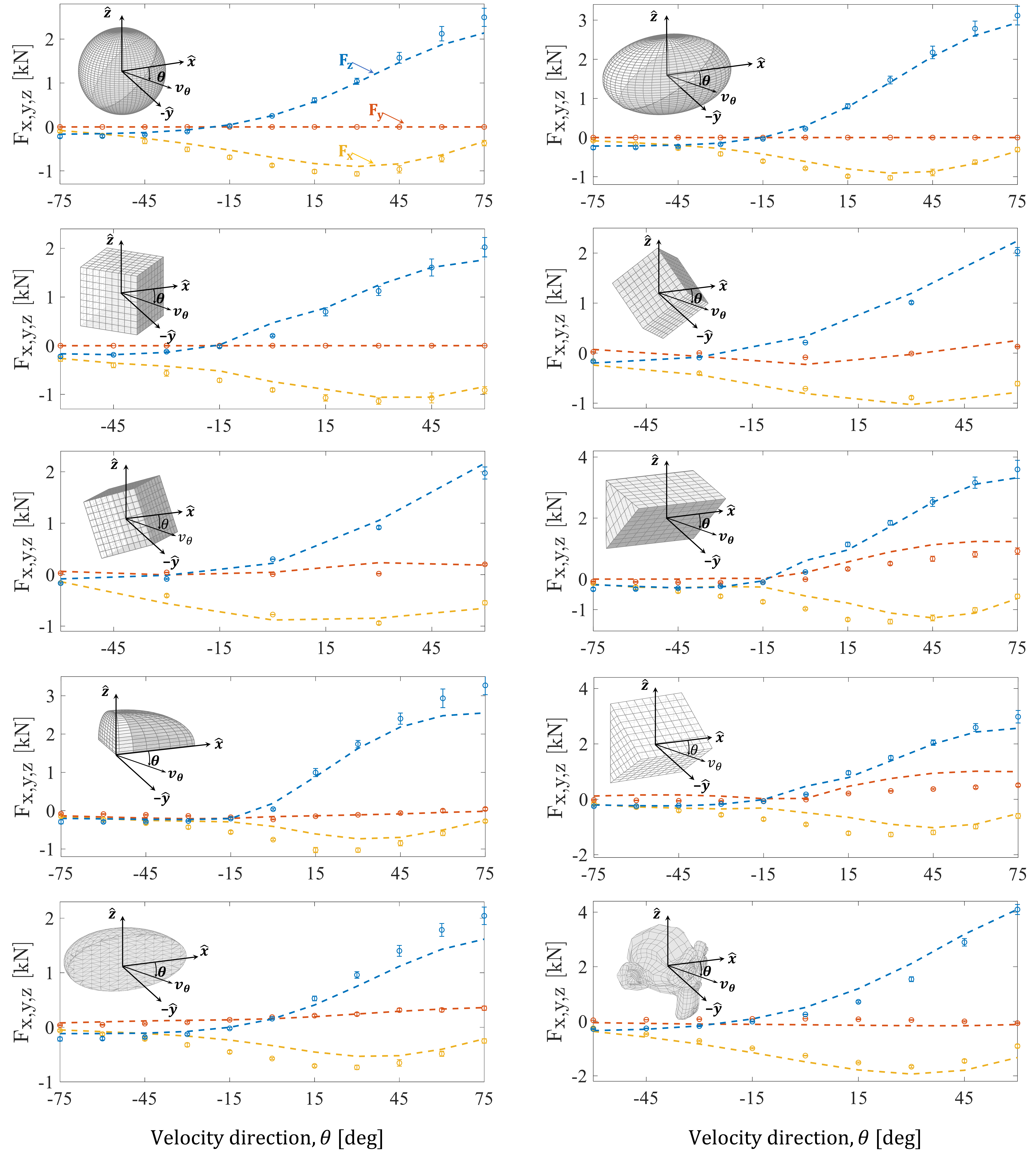}
\caption{\emph{3D-RFT Verification Studies 1-10}: Variation of different force components ($F_x$: yellow, $F_y$: orange, and $F_z$: blue) during motions of various rigid objects (intruders) obtained from continuum modeling (`o' markers) and 3D-RFT (dashed lines) at various velocity directions ($\hat{\boldsymbol{v}}_\theta$). All the studies were conducted at a velocity magnitude of $0.1$ m/s. $\theta$ represents the angle between $\hat{\boldsymbol{v}}_\theta$ and the positive x-axis. All the velocities completely lie in the $xz-$plane. A pictorial representation of each intruder is provided in the corresponding sub-figure. The intruder shapes include \emph{(1)} a $5$ cm radius sphere, \textit{(2)} an ellipsoid with $[7.5,4.5,4.5]$ cm semi-axes (x,y,z), \textit{(3)} a $7.5$ cm tilted cube, rotated from a cartesian alignment by $\pi/4$ radians about the $z$-axis, \textit{(4)} a $7.5$ cm cube sequentially rotated by $\pi/3$ and $\pi/4$ radians along the $y$-axis and $z$-axis from a cartesian alignment, \textit{(5)} a $7.5$ cm cube sequentially rotated by $\pi/6$ and $\pi/3$ radians  along the $y$-axis and $z$-axis from a cartesian coordinate alignment,  \textit{(6)} an isosceles right angle prism with $7.5$ cm equal sides and $10.5$ cm width, \textit{(7)} a quarter ellipsoid with $[7.5,4.5,4.5]$ cm semi-axes (x,y,z) ($x>0$ and $y>0$), \textit{(7)} an iscosceles right angle prism with equal sides of $10.5$ cm and $7.5$ cm width, \textit{(9)} a half-ellipsoid with $[7.5,4.5,4.5]$ cm semi-axes (x,y,z) ($y>0$), and \textit{(10)} a monkey head shape from the open-source 3D computer graphics software \textit{`Blender'} at a scale factor of $0.075$ and facing $\pi/4$ radians from the positive $x$-direction  in the $xy$-plane.% The continuum modeling simulations use an effective material density ( $\rho_c = \phi_c \times \rho_g$) of $3000$ kg/m$^3$, an internal friction ($\mu_{\textrm{int}}$) of $0.4$, and surface friction ($\mu_{\textrm{surf}}$) of $0.4$. 
%All the objects were submerged to an initial center-depth of $27$ cm. 
}
\label{fig:verification_all}
\end{figure*}

Figure \ref{fig:3drftgraphs_and_setup}E shows a comparison of reference data to an example fitting of 3D-RFT. Odd columns in the figure show the data obtained using continuum simulations as a function of $\beta$ and $\gamma$ at four $\psi$ values. The material properties were $\mu_{\textrm{int}}=0.4$, $\rho_c=3000$ kg/m$^3$, and $\mu_{\textrm{surf}}=0.15$. Corresponding 3D-RFT fittings are plotted on the even columns. We find the value of the scaling coefficient $\xi_n$ to be $0.92\times10^6$ N/m$^3$ for this material. While Eq \ref{3drft_eq} represents the most generic form of 3D-RFT, the choice of the functions $f_i$ determines the final 3D-RFT model. All the results presented in this work use $3^{\textrm{rd}}$ degree polynomial fits for the $f_i$ functions (Table S3).
%for a total of 20 coefficients.  This should be compared to the 10 coeficients typically used in 2D-RFT \cite{li2013terradynamics}; the physical constraints imposed prevented exponentially more fitting from being needed in 3D.  
Higher-order polynomials could be used, which can better fit the reference data. We provide one such form in the Supplementary Information (Table S4). The performance of 3D-RFT does not change significantly between $3^{\textrm{rd}}$ and $4^{\textrm{th}}$ degree polynomial fits. The latter form fits the trends of $|\boldsymbol{\alpha}_t|/|\boldsymbol{\alpha}_n|$ better but has inconsequential effects on 3D-RFT predictions for the test cases used in this study. 

The 3D-RFT model we propose is completed using cubic $\hat{f}$ fit as shown in Fig S4 --- this dependence is in accord with observations of past researchers in the simpler vertical intrusion of flat plates \cite{kang2018archimedes} --- and with $\boldsymbol{\alpha}^{\text{gen}}$ expressed using Eqs \ref{3drft_eq} in terms of third degree polynomial fits for the $f_i$, and using directions $\{\hat{\boldsymbol{r}},\hat{\boldsymbol{\theta}},\hat{\boldsymbol{z}}\}$ and angles $\{\beta, \gamma, \psi\}$ as shown in Fig \ref{fig:3drftgraphs_and_setup}A-D.  To numerically implement the model, we discretize the intruder surface into small plate elements and determine $\{\beta, \gamma, \psi\}$ and $\{\hat{\boldsymbol{r}},\hat{\boldsymbol{\theta}}\}$ for each element.  The model then provides the force on each element that is on the leading edge of the intruder. A step-by-step implementation strategy for 3D-RFT is given in Sec S7 of the Supplementary Information.

\begin{figure*}[ht!]
\centering
\includegraphics[trim = 0mm 0mm 125mm 0mm, clip, width=1.0 \linewidth] {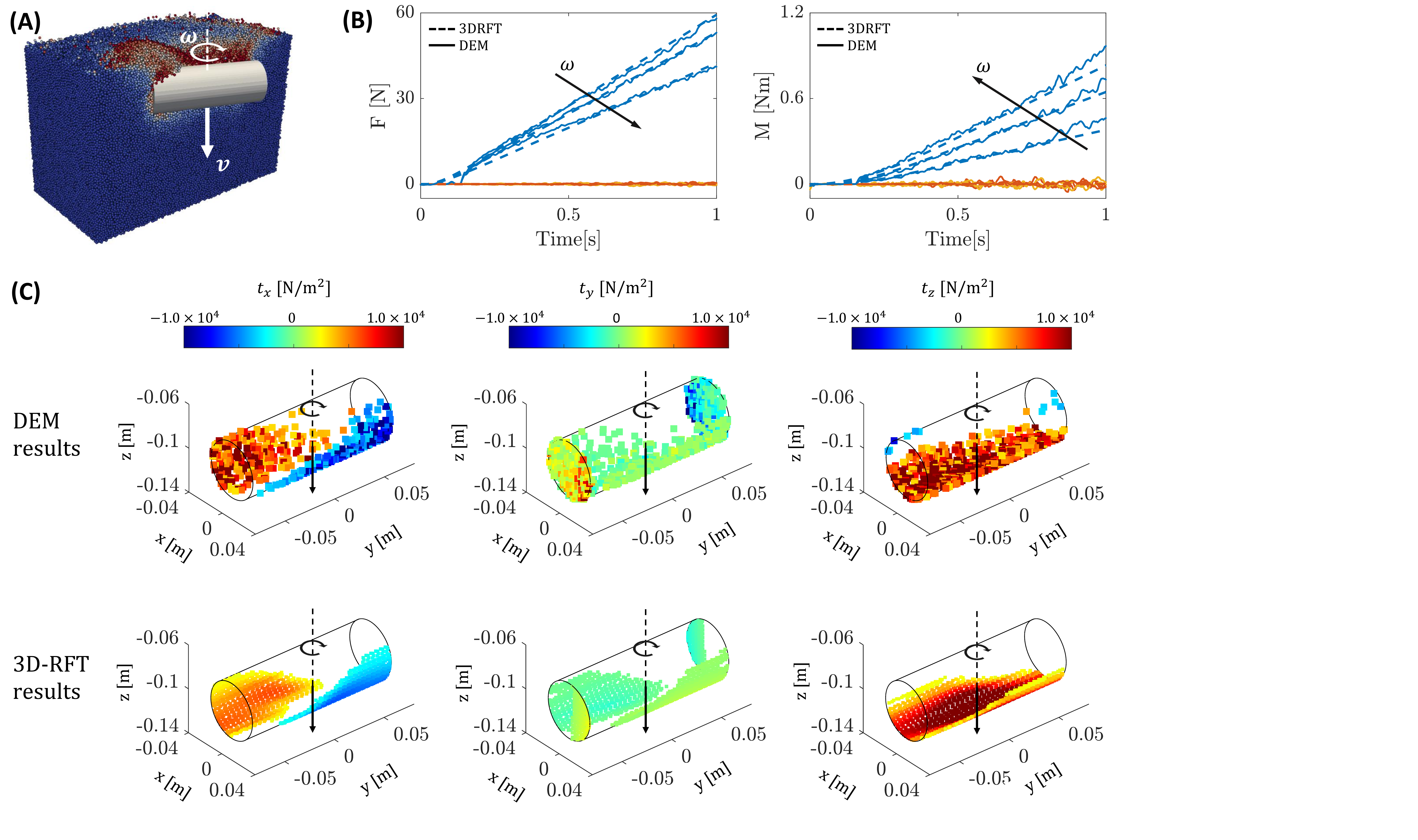}
\caption{\emph{DEM based 3D-RFT verification: Cylinder drill}: \emph{(A)} A snapshot of the cylinder drill setup where a $50$ mm diameter and $140$ mm length cylinder was simultaneously {rotated ($\omega$, clockwise) and translated ($v$, downwards) along the $z$-axis. We use three combinations of $(\omega, v )$: $[(0.25\pi,0.1), ( 0.5\pi,0.1), (\pi,0.1)] $ (rad/s,m/s). Black arrows in \emph{(B)} show the direction of increasing $\omega$ in each graph}. The grains are colored with velocity magnitudes. The simulation domain consisted $\sim6\times10^5$ particles ($50/50$ mix of $3$ mm and $3.4$ mm diameter ($d$) grains) spread over $100d\times100d\times70d$ physical space. \emph{(B)} Variation of net force ($F$, left) and moment ($M$, right) components ($x$: yellow, $y$: orange, and $z$: blue) from DEM (solid lines) and 3D-RFT (dotted lines) for $\omega= \pi$ rad/s . \emph{(C)} Variation of various force components from DEM (Top) and 3D-RFT (Bottom) at a $10$ cm depth below the free surface ($t=1s$). The DEM material properties are provided in Table S2 of the Supplementary Information.}
\label{fig:cylinder_drill}
\end{figure*}

\begin{figure*}[ht!]
\centering
\includegraphics[trim = 0mm 0mm 40mm 0mm, clip, width=1.0 \linewidth] {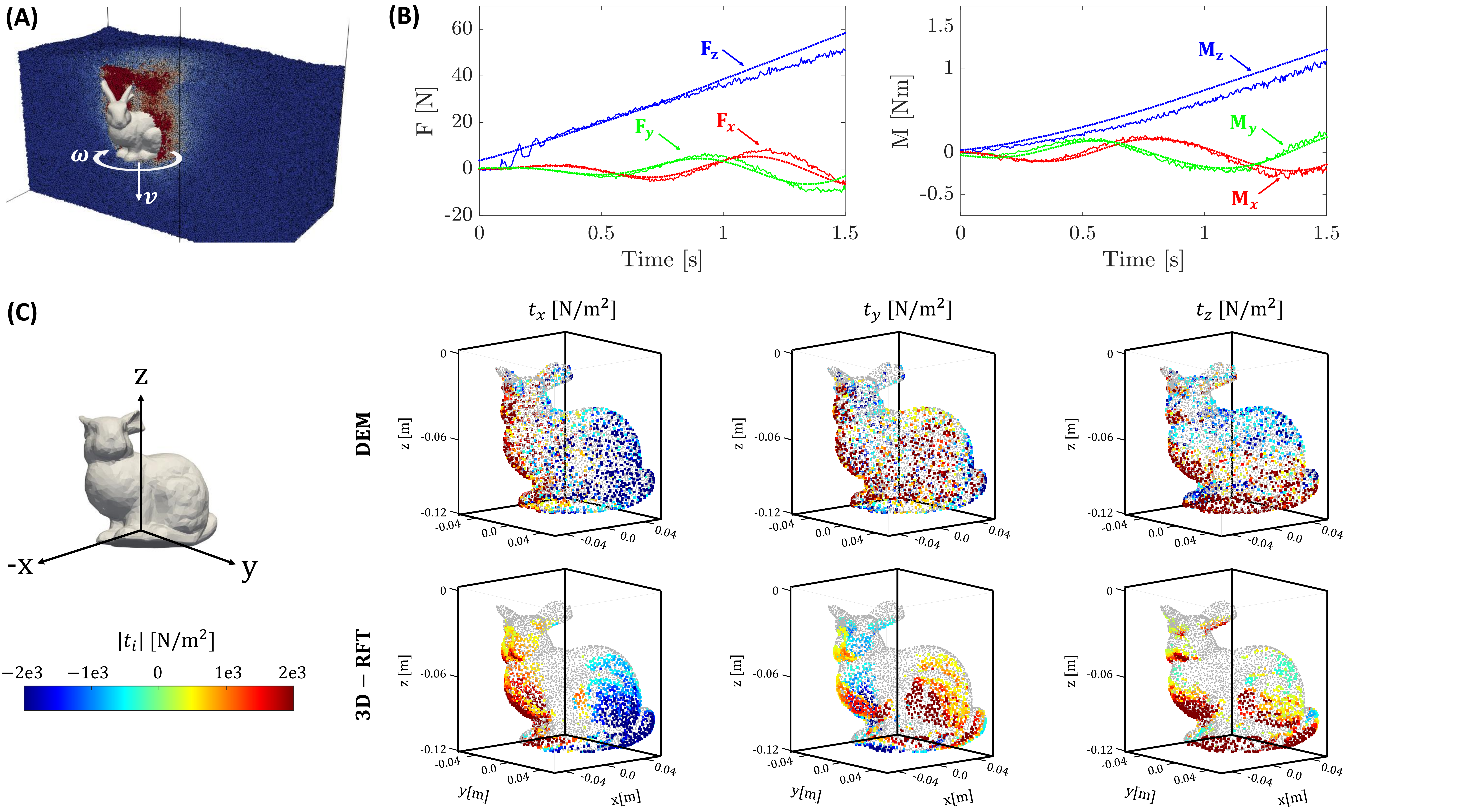}
\caption{\emph{DEM based 3D-RFT verification: Bunny drill}:  \emph{(A)} A snapshot of the Stanford-Bunny drill setup where a $10$ cm high stanford-bunny was simultaneously rotated ($\omega= 2\pi $rad/s, clockwise) and translated ($v=0.1$m/s, downwards) along the $z$-axis. The grains are colored with velocity magnitudes. The simulation domain consisted of $\sim2.1\times10^6$ particles ($50/50$ split of $3$ mm and $3.4$ mm diameter ($d$) grains) spread over a $150d\times150d\times88d$ physical space. \emph{(B)} Variation of net force ($F$, left) and moment ($M$, right) components ($x$: yellow, $y$: orange, and $z$: blue) from DEM (solid lines) and 3D-RFT (dotted lines). \emph{(C)} Commponets of the surface stress distribution from DEM (Top) and 3D-RFT (Bottom) at a $5$ cm bunny-center-depth below the free surface. The DEM material properties are provided in Table S2 of the Supplementary Information.}
\label{fig:bunny_drill}
\end{figure*}

\section{Validation studies}
We first test the accuracy of the implied localization of the proposed form of 3D-RFT (Eq \ref{eq:genericrft}) as well as the $f_i$ fits by comparing  predictions for ten arbitrary intruding objects to full continuum model solutions of the same intrusions. We use the continuum material properties $\mu_{\textrm{int}}=0.4$, $\rho_c=3000$ kg/m$^3$, and $\mu_{\textrm{surf}}=0.4$ for these cases.  %We use a 3D-RFT scaling coefficient ($\xi_n$) value of $0.92\times 10^6$ N/m$^3$ based on the $\hat{f}$ relationship between $\xi_n/\rho_c g$ and $\mu_{\text{int}}$ shown in Fig S3. See Table S1 and section S2 for more details. 
A representation of the objects and their dimensions are provided in Fig \ref{fig:verification_all} and its caption. The object length scales are kept to be $7$ cm in all the cases, and the objects are submerged to an initial depth of $27$ cm (vertical distance between the free surface and the geometric center of the shape). The objects are moved at a speed of $0.1$ m/s in different directions in the $xz$-plane. These directions are characterized using $\theta$, which represents the angle between the velocity direction ($\hat{\boldsymbol{v}}_{\theta}$) and the positive $x$-axis in a clockwise direction (same as $\gamma$ definition for a plate element). Negative $\theta$ represents upward motion,   positive $\theta$ represents downward motion, and $\theta=0$ represents horizontal motion along the $x$-direction. The variations of net-force ($F_x$, $F_y$, and $F_z$) with $\theta$ are plotted in Fig \ref{fig:verification_all}. 3D-RFT agrees with the continuum solutions well in modeling all the intrusion test scenarios considered in Fig \ref{fig:verification_all}.  Objects with sharp corners generally show somewhat weaker fits than those with smoother shapes; this could be because sharp corners are difficult to represent with our material point method.  

%We test the accuracy of 3D-RFT in modeling motion of ten rigid objects shown in figure \ref{fig:verification_all} in various directions. The rigid objects include \emph{(1)}a $5$ cm radius sphere, \emph{(2)} an ellipsoid with $[7.5,4.5,4.5]$ cm semi-axes(x,y,z), \emph{(3)} a $7.5$ cm edge cube rotated by $\pi/4$ radian along $z$-axis passing through its center from the initial alignment to cartesian coordinate frame, \emph{(4)} a $7.5$ cm cube first rotated by $\pi/3$ radian along $y$-axis and then rotated by $\pi/4$ along $z$-axis passing through its center from the initial alignment to cartesian coordinate frame, \emph{(5)} an isosceles right angle prism with equal sides of $7.5$ cm and $10.5$ cm width,\emph{(6)} a isosceles right angle prism with equal sides of $10.5$ cm and $7.5$ cm width, \emph{(7)} an quarter ellipsoid with $[7.5,4.5,4.5]$ cm semi-axes(x,y,z) ($x>0$ and $y>0$), \emph{(8)} a $7.5$ cm cube first rotated by $\pi/6$ radian along $y$-axis and then rotated by $\pi/3$ along $z$-axis passing through its center from the initial alignment to cartesian coordinate frame,\emph{(9)} an half-ellipsoid with $[7.5,4.5,4.5]$ cm semi-axes(x,y,z) ($y>0$), and \emph{(9)} a monkey shape from \textit{'Blender'} (a free and open-source 3D computer graphics software) facing $\pi/4$ rad from x-direction. 

\subsection*{Validation of 3D-RFT with detailed DEM studies}
We further check the performance of 3D-RFT with two DEM studies. In these studies, we measure net moment, net force, and resistive force distribution on bodies intruding into granular volumes with simultaneous rotation and translation velocities. We use a 50/50 mixture of $3$ mm and $3.4$ mm diameter $(d)$ grains with a grain density of $2470$ kg/m$^3$ and the  granular volumes have an effective bulk density of $1310$ kg/m$^3$ ($\phi \approx 0.53$) in both the DEM studies. We determine the internal coefficient of friction $\mu_{\text{int}}$ as $0.21$ using simple shear simulations. Section S8 of the Supplementary Information provides more details of the simple shear test setup and detailed material properties. Using this value together with the known $\hat{f}$ relationship, We obtain a scaling coefficient ($\xi_n$) value of $0.12\times 10^6$ N/m$^3$. See Table S1 and Sec S4 for more details.\\

\noindent \textit{Cylinder Drill:} In this test, we model simultaneous rotation and translation (drilling) of a \textit{solid cylindrical} intruder  along the $z$-axis \textcolor{black}{(vertically down)} in a granular volume {(diameter$=0.05$ m, length=$0.14$ m)}. \textcolor{black}{The cylinder axis was kept in the horizontal plane throughout the motion}. The setup consists of approximately $6\times10^5$ particles in a $100d\times100d\times70d$ sized granular bed. The setup dimensions and setup schematic are provided in Fig \ref{fig:cylinder_drill}. The figure also shows the variations of force and moment on the intruder over time from the DEM studies versus 3D-RFT. In addition, the figure shows the variation of stress over the intruder surface from DEM and 3D-RFT. All reported components (net force and moments, as well as stress distributions) show a strong match between the two approaches.\\

\noindent\textit{Bunny Drill:}  In this test, we model the drilling motion $(\omega=2\pi\,\text{rad/s}, v=0.1 \, \text{m/s} )$ of a \textit{Stanford Bunny} \cite{turk1994zippered} shaped rigid intruder in a granular volume. {The shape is chosen because it is an example of a complex, asymmetric 3D object.} The granular bed consists of approximately $2.1\times 10^6$ particles spread over a $150d\times150d\times88d$ sized domain. The bunny shape was slightly modified from the standard shape --- the shape was proportionally scaled in such a way that the bunny height measures $0.1$ m, and the bunny base was flattened to make the base a plane surface without an inward extrusion. Figure \ref{fig:bunny_drill} shows the simulation setup where the grains are colored with velocity magnitudes. Figure \ref{fig:bunny_drill} also shows the variation of stresses over the intruder surface from DEM and 3D-RFT. All the reported components (net force and moments, as well as stress distributions) show a strong match between the two approaches. 

\section{Conclusion} 
{This work proposes a mechanistic framework for developing reduced-order models in soft-materials. Successful development of a granular 3D-RFT that overcomes the limitations of previous attempts in this direction (see Sec 1 of Supplemental 27
Information) indicates the robustness of the approach for these purposes}. The 3D-RFT developed herein is an important step towards developing a generic real-time modeling technique capable of modeling granular intrusion of arbitrarily shaped objects over a large range of low and high-speed scenarios in diverse materials and environments. Previously, granular RFT's usage has focused on the modeling of arbitrary 2D objects moving in-plane. We have proposed an extension of RFT to three dimensions in a fashion consistent with granular continuum mechanics and necessary symmetry constraints. The accuracy of the proposed 3D-RFT was demonstrated against a variety of full-field intrusion simulations, both continuum and DEM. Notably, we provide a scheme that determines 3D-RFT in different intrusion systems quickly and directly in terms of basic properties of the granular media ($\rho_c$ and $\mu_{\text{int}}$) and the intruder surface ($\mu_{\text{surf}}$).  The most immediate opportunity to expand 3D-RFT would be to combine 3D-RFT with Dynamic RFT \cite{agarwal2021surprising} to build a high-speed three-dimensional RFT (3D-DRFT). { The current form of 3D-RFT does not include a ``shadowing effect'' i.e. the fact that forces are reduced on leading edge surfaces that lie in the immediate wake behind another part of the intruder \cite{suzuki2019study}. Such effects are more pronounced in intruders with complex shapes or fine geometric features such as the Bunny shape we consider in this study. Characterizing this effect would be an important addition to RFT}. Effects of multibody intrusions \cite{pravin2021effect,agarwal2021efficacy}, density variations \cite{gravish2010force}, inertial and non-inertial velocity effects \cite{katsuragi2007unified,schiebel2020mitigating,agarwal2021surprising}, cohesion \cite{francoeur2014burrowing,athani2021pulling}, and inclined domains \cite{humeau2019locomotion} on the resistive forces experienced by intruding bodies are among other aspects for further exploration toward the ultimate goal of a generic and fast granular intrusion model {applicable to terradynamical motions \cite{he2019review},  granular impact systems \cite{soliman1976effect,wright2020ricochets}, locomotors  \cite{astley2020surprising}, and many other similar applications}.

\textcolor{black}{The three-step mechanistic approach that produced 3D-RFT could be extended to produce intrusion models in other soft materials, including possible applications in colloids, gels, and biological media. We refer readers to Sec S2 of the Supplementary Information to see demonstrations of the procedure being used in other common materials and a discussion on determining the accuracy of the RFT localization rule (Step 1) in other media.
%Though the final functional thus produced could be different than RFT, the basic mechanistic steps remain the same. Granular RFT is just an example of a reduced-order model which derives itself with this approach and experimental observations of stress-localization and hydrostatic variation of material resistance. Other materials which may not follow RFT-like constraints could still use our approach. For such materials, Step 1 would still be the identification of a simple continuum representation of the system with a minimal set of system properties in such a way that the phenomenons of interest are appropriately represented. In recent years, many researchers have proposed continuum theories for a variety of soft materials such as colloids and biological matters \cite{kosmrlj2011continuum, byrne2009individual}. Such models could directly be used in Step 1 of reduced-order model development. The scaling and dimensional arguments will remain the same for developing a valid reduced order form in Step 2. Subject to the similarity of the system representation, a plate for 3D-RFT, the 3DRFT universal constraints we use could be directly used for any material in Step 3. 
} 

\matmethods{
\subsection*{Evaluation of quasi-static conditions in a system } \label{quasi-static-numbers} 
 We use the following definitions of the micro-inertial number $I$ and the macro-inertial number $I_{\text{mac}}$ for evaluating the applicability of 3D-RFT in modeling the granular resistive forces in a granular intrusion system;
\begin{align}
& I=\dot{\gamma}/\sqrt{P/\rho_g d^2}\ , \qquad \qquad I_{\text{mac}} = v/\sqrt{P/\rho_g}\, , \nonumber
\end{align}
where $\dot{\gamma}$ represents the material shear rate, $P$ represents the hydro-static pressure, $\rho_g$ represents the material grain density, $d$ represents the mean grain diameter, and $v$ represents the speed. The $I_{\text{mac}}$ formulation is equivalent to the inverse square root of the \textit{Euler number} which measures the ratio of the dynamic pressure $\rho v^2$ to the total pressure $P$. 

The macro- and micro-inertial numbers are defined pointwise within a granular media so, to determine if an intrusion is quasi-static, it is convenient to determine characteristic values for these numbers.  For this, we use characteristic values of $\dot\gamma, \, P, \, \text{and} \, v$ . We assume that the intruder has an angular velocity $\omega$, a translational velocity $v_\text{intru}$, and a characteristic length $L$. We also assume that the media has a critical density $\rho_c$ and that the system is acted upon by gravity $g$. We characterize $v$ as $v=\text{max}(v_{\text{intru}}, L\omega/2)$ and $\dot\gamma$ as $\dot{\gamma} = v/L$. We consider intrusive loading of the system at characteristic depth $L$ to give a characteristic $P$ as $\xi_n L$.  
Upon substitution, we get:
\begin{align}
&I \sim (v/L)/\sqrt{(\xi_n L)/\rho_g d^2} = \sqrt{\frac{v^2 \rho_g d^2}{\xi_n L^3}} = \frac{vd}{L}\sqrt{\frac{\rho_g}{\xi_n L}}\, , \nonumber \\
&I_\text{mac} \sim v/\sqrt{\xi_n L/\rho_g} = v\sqrt{\frac{\rho_g}{\xi_nL}}\, .
\nonumber 
\end{align}
From the above equations, we observe that the characteristic value of $I_\text{mac}$ reduces to a multiple of the \textit{Froude number} $(Fr)$ in gravity loaded systems. To this end, Sunday et al. \cite{sunday2022influence} explored the existence of macro-inertial effects during high-speed granular intrusions and observed insignificant contributions of macro-inertial effects in the material force response for $Fr<1.5$, which gives $I_{\text{mac}}\lessapprox 0.48 $. {Similarly, Agarwal et al. \cite{agarwal2021surprising}, observed insignificant macro-inertial effects (macro-inertial forces < 10\% of static resistive forces i.e. $\rho A v^2/K|z|<10\%$) in granular plate intrusions at $I_\text{mac} < 0.16$.  
% v_cutoff = \sqrt((Kz/rhoA)*10%) = \sqrt((580*0.03*.10)/(900*0.016)) = 0.25, => I_mac= 0.25\sqrt(900 / (31 rho g) *0.016) = 0.25\sqrt{1/31*10*0.016}=0.35*0.44 = 0.16
Thus, we impose an upper limit of $0.15$ on $I_\text{mac}$ to be quasi-static. The amount $I$ affects the flow can be quantified by how much it changes the apparent internal friction \cite{gdr2004dense,da2005rheophysics}. To keep these changes bounded by 10\% we set an upper bound on the characteristic value of $I$ to be $0.010$ so as to ensure quasi-static conditions. %In brief, keeping $I$ low prevents the resistive force contributions from collision character of granular flows \cite{gdr2004dense} and the keeping $I_\text{mac}$ low prevents any significant macro-inertial contributions to granular force response upon intrusion. Both of these effects can not be captured by proposed 3D-RFT form and are thus avoided by setting limits on $I$ and $I_\text{mac}$. 
For all the test cases used in this study, we choose system parameters in such a way that $I$ and $I_{\text{mac}}$ are always below above mentioned limits keeping their motions in quasi-static limits. The test cases 1-10 are continuum simulation that use a rate-independent constitutive law and have $I_{\text{mac}} \sim 0.02$ $( L\approx0.07\, \text{m},\, \,v=0.1\, \text{m/s}, \, \xi_n = 0.92\times 10^6, \, \rho_c = 3000 \text{kg/m}^3)$. In the DEM based cylinder drill test cases, we find $I < 0.002$ and $I_\text{mac} < 0.07$ $( L\approx0.10\, \text{m},\, \omega<\pi\, \text{rad/s}, \,v_{\text{intruder}}=0.1\, \text{m/s}, \, \xi_n = 0.12\times 10^6, \, \rho_g = 2470 \text{kg/m}^3)$. Similarly, in the bunny drill test case, we find $I \approx 0.004$ and $I_{\text{mac}} \sim 0.13$ $( L\approx0.10\, \text{m},\, \omega=2\pi\, \text{rad/s}, \,v_{\text{intruder}}=0.1\, \text{m/s}, \, \xi_n = 0.12\times 10^6, \, \rho_g = 2470 \text{kg/m}^3)$. Thus, 3D-RFT is a valid approach for modeling all the test cases considered in this study, based on the insignificance of micro- and macro- inertial force contributions}. 

\subsection*{Continuum approach accuracy validation}
Several studies in the past have verified the accuracy of this constitutive formulation in plane-strain problems(2D). We use the 3D numerical implementation of MPM developed by Baumgarten and Kamrin \cite{baumgarten2019general} for this study which has been successfully used for modeling complex problems in the past \cite{baumgarten2019general,baumgarten2019general2}. For the continuum model to be useful to determine input data for a 3D-RFT, it must be shown to reliably match experiments for 3D plate intrusions. We test this in two scenarios.

% \begin{figure}[ht!]
% \centering
% \includegraphics[trim = 0mm 0mm 15mm 0mm, clip, width=1.0 \linewidth] {graphics/exp_Li_etal.pdf}
% \caption{\emph{Experiment vs Continuum Model --- Comparison of in-plane plate motions}: \emph{(A)}Schematic of plate orientation angle $\beta$ and $\gamma$ for in-plane motion study conducted using 3D  simulation setup shown in figure \ref{fig:data_collection}. \emph{(B)}Force/area/depth ($\boldsymbol{\alpha}$) from Li et al.\cite{li2013terradynamics} experiments (top) and continuum simulations (bottom). The plate configurations are also overlayed on graphs for clarity. The plates had no twist ($\psi=0$) in regards to 3D-RFT definitions in these tests. Both the experiments and the simulations use glass beads with grain density ($\rho_g$) of $2500$ kg/m$^3$ and a packing farction ($\phi_c$) of $0.58$. Internal friction is $\mu_\textrm{int} = 0.4$ and surface friction, $\mu_{\textrm{surf}} =0.4$ for continuum simulations to match reported values in Li et al.\cite{li2013terradynamics}. %The colorbar limits are slightly different between the experiments and the simulations because we do not attempt exact property calibration ($\mu_{\textrm{int}}$ and $\mu_{\textrm{surf}}$) for this test case although we keep them in same the practically expected range.}
% }
% \label{fig:exp_sim_2drft}
% \end{figure}

In the first test case, we check if the 3D-continuum simulations can regenerate the experimental variation of force/depth/area on flat plates in submerged granular beds from Li et al.\cite{li2013terradynamics}. This experimental data was also used by Li et al.\cite{li2013terradynamics} in the generation of 2D-RFT form.  We use an effective materials density of $\rho_c=1450$ kg/m$^3$ (loose glass beads, $\rho_g=2500$ kg/m$^3$, $\phi_c=0.58$) inline with Li et al.\cite{li2013terradynamics} experiments and an approximate internal friction value for glass beads as $\mu_{\textrm{int}} = 0.4$. The  media-plate surface friction was taken as $\mu_{\textrm{surf}}=0.4$. The relative values of the forces from continuum results remarkably match the experimental observations. The absolute values from continuum results, however, are higher than experiments by a constant multiplicative factor of $\sim 1.1$. A smaller value of $\mu_{\textrm{int}}$ for glass beads could have provided a closer match to the experiments as the graphs are not expected to change their shape with changing internal friction values \cite{li2013terradynamics}. But we do not attempt the exact calibration as the purpose of the test was to verify the accuracy of the continuum formulation and implementation. These results establish sufficient efficacy of the continuum model for plate motions in which the velocity, plate normal, and gravity or co-planar. %. Additionally as $0.4$. As Li et al.\cite{li2013terradynamics} experiments also reported that the shape of these graphs does not vary with the choice of non-cohesive granular material. The continuum modeling also gave out-of-plane force components. We found to be insignificant ($<0.001 N/cm^3$) for all combinations of $\beta$ and $\gamma$ (see Fig \ref{fig:2drftgraphs}A for angles definitions) and do not plot them.

In the second test case, we assess the quantitative accuracy of the continuum approach in modeling in-plane as well as out-of-plane forces. We consider a study Maladen et al.\cite{maladen2011mechanical} which measured the normal and tangential forces on submerged plates moving horizontally in granular media as a function of plate twist (see Fig \ref{fig:exp_maladen_etal} (top) for angles definition). The material properties are provided in the figure caption. The continuum results match observations from Maladen et al.\cite{maladen2011mechanical} well.

The combination of the above two studies establishes the overall accuracy of the continuum model and its implementation for both in-plane and out-of-plane inputs and outputs in plate intrusion problems. %Thus, we use continuum modeling results as the reference solutions for the rest of the study.

% \begin{figure}[ht!]
% \centering
% \includegraphics[trim = 0mm 10mm 180mm 0mm, clip, width=1.0 \linewidth] {graphics/exp_maladen_etal.pdf}
% \caption{
% \emph{Experiment vs Continuum Model: Dependence on twist angle}: (A) Schematic of plate orientations, and (B) variations of normal (red) and tangential (blue) forces from Maladen et al \cite{maladen2011mechanical} experiments ($\bullet$ marker), their analytical fits to their results (dotted lines), and continuum simulations ($\blacksquare$ marker with solid line). The forces are normalized by the plate center-depth ($|\boldsymbol{z}|$) and plate area. Experiments (loosely packed $3$ mm glass particles) as well as simulation use glass beads ($\rho_g = 2500$ kg/m$^3$ and $\rho_c=0.6$) as the granular media. Continuum simulations use $\mu_{\textrm{int}}=0.4$ and $\mu_{\text{surf}}=0.27$ in accordance with reported experimental values. The original Maladen et al. \cite{maladen2011mechanical} results used a twist angle ($\beta_d = \pi/2-\psi$) as the $x$-axis in their plots. We have modified the plots to have $\psi=\pi/2-\beta_d$ on $x$-axis for simplifying the discussion.}
% \label{fig:exp_maladen_etal}
% \end{figure}

\begin{figure*}[ht!]
\centering
\includegraphics[trim = 0mm 0mm 0mm 0mm, clip, width=1.0 \linewidth] {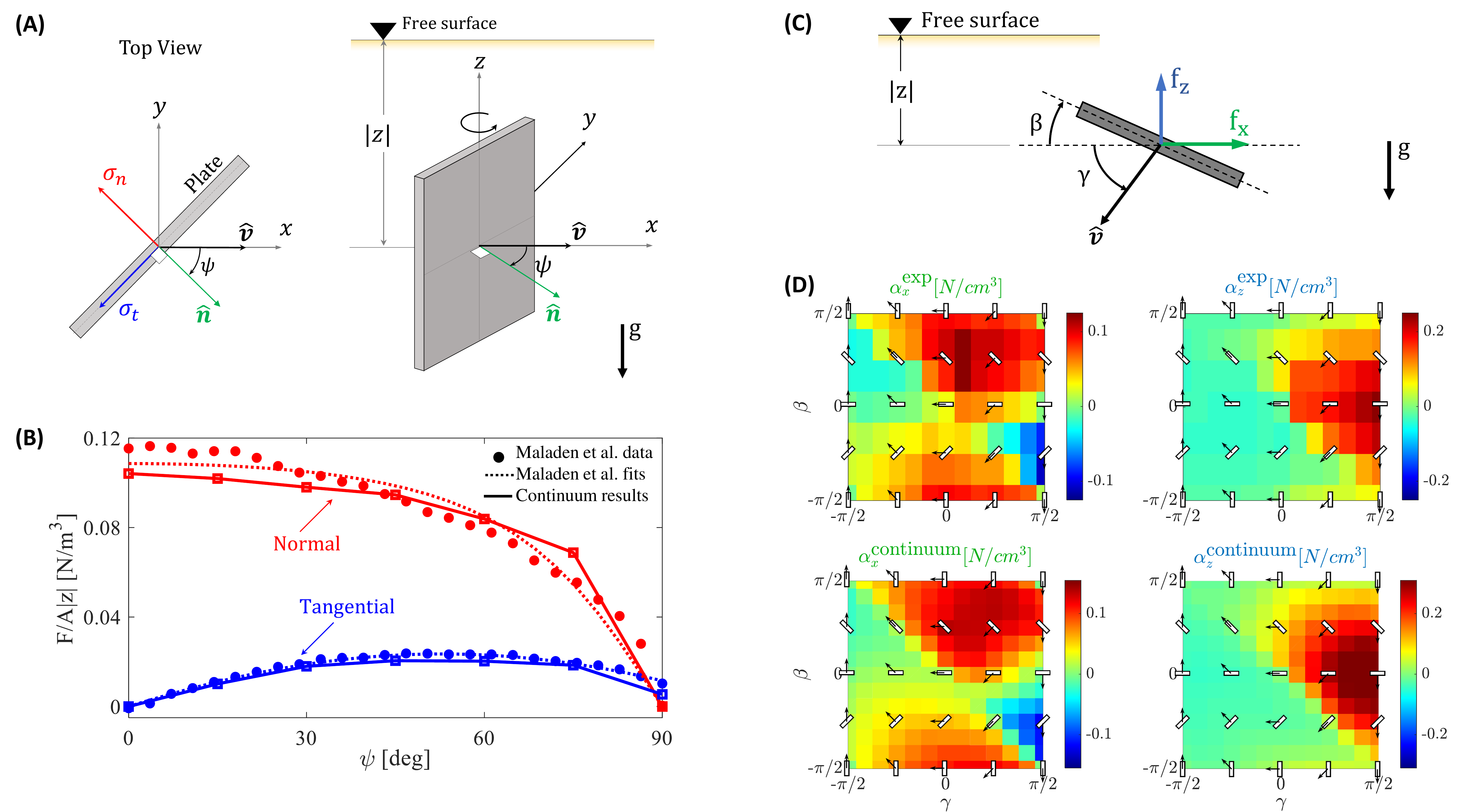}
\caption{
\emph{Experiment vs Continuum Model --- Dependence on twist angle}: (A) Schematic of plate orientations, and (B) variations of normal (red) and tangential (blue) forces from Maladen et al \cite{maladen2011mechanical} experiments ($\bullet$ marker), their analytical fits to their results (dotted lines), and continuum simulations ($\blacksquare$ marker with solid line). The forces are normalized by the plate center-depth ($|\boldsymbol{z}|$) and plate area. Experiments (loosely packed $3$ mm glass particles) as well as simulation use glass beads ($\rho_g = 2500$ kg/m$^3$ and $\rho_c=0.6$) as the granular media. Continuum simulations use $\mu_{\textrm{int}}=0.4$ and $\mu_{\text{surf}}=0.27$ in accordance with reported experimental values. The original Maladen et al. \cite{maladen2011mechanical} results used a twist angle ($\beta_d = \pi/2-\psi$) as the $x$-axis in their plots. We have modified the plots to have $\psi=\pi/2-\beta_d$ on $x$-axis for simplifying the discussion. --- \emph{Comparison of in-plane plate motions}: \emph{(C)} Schematic of plate orientation angle $\beta$ and $\gamma$ for in-plane motion study conducted using 3D  simulation setup shown in figure \ref{fig:3drftgraphs_and_setup}A-D. \emph{(D)}Force/area/depth ($\boldsymbol{\alpha}$) from Li et al.\cite{li2013terradynamics} experiments (top) and continuum simulations (bottom). The plate configurations are also overlayed on graphs for clarity. The plates had no twist ($\psi=0$) in regards to 3D-RFT definitions in these tests. Both the experiments and the simulations use glass beads with grain density ($\rho_g$) of $2500$ kg/m$^3$ and a packing farction ($\phi_c$) of $0.58$. Internal friction is $\mu_\textrm{int} = 0.4$ and surface friction, $\mu_{\textrm{surf}} =0.4$ for continuum simulations to match reported values in Li et al.\cite{li2013terradynamics}. 
}
\label{fig:exp_maladen_etal}
\end{figure*}

}
\showmatmethods{} % Display the Materials and Methods section

\acknow{SA, and KK acknowledge support from Army Research Office (ARO) grants {W911NF1510196 and W911NF1810118}, support from the U.S. Army Tank-Automotive Research, Development and Engineering Center (TARDEC), and NASA STTR Award Number 80NSSC20C0252. DG acknowledges support from ARO grant W911NF-18-1-0120. SA thanks Aaron Baumgarten, who provided the authors with his 3D-MPM model implementation. We also thank Andras Karsai for helpful discussions on potential forms of 3D-RFT. Portions of the paper were developed from the PhD thesis of SA \cite{agarwal2022PhD}.}

\showacknow{} % Display the acknowledgments section

\bibliography{pnas-sample}

% \section*{.}
\newpage

\end{document}